Research

**Cite this article:** Foulds S, Prove O, Kendon V. 2024 Generalizing multipartite concentratable entanglement for practical applications: mixed, qudit and optical states. *Phil. Trans. R. Soc. A* **382**: 20240411.
https://doi.org/10.1098/rsta.2024.0411

Received: 13 August 2024
Accepted: 18 November 2024

One contribution of 18 to a theme issue 'The quantum theory of light'.

**Subject Areas:**
optics, quantum physics

**Author for correspondence:**
Steph Foulds
e-mail: s.foulds@strath.ac.uk

# Generalizing multipartite concentratable entanglement for practical applications: mixed, qudit and optical states

Steph Foulds[1,2], Oliver Prove[2] and Viv Kendon[1,2]

[1]Physics Department, University of Strathclyde, Glasgow G4 0NG, UK
[2]Physics Department, Durham University, South Road, Durham DH1 3LE, UK

SF, 0000-0001-9714-2184

The controlled SWAP test for detecting and quantifying entanglement applied to pure qubit states is robust to small errors in the states and efficient for large multi-qubit states (Foulds *et al*. 2021 *Quantum Sci. Technol*. **6**, 035002 (doi:10.1088/2058-9565/abe458)). We extend this, and the related measure *concentratable entanglement* (CE), to enable important practical applications in quantum information processing. We investigate the lower bound of concentratable entanglement given in (Beckey *et al*. 2023 *Phys. Rev. A* **107**, 062425 (doi:10.1103/physreva.107.062425)) and conjecture an upper bound of the mixed-state concentratable entanglement that is robust to c-SWAP test errors. Since experimental states are always slightly mixed, our work makes the c-SWAP test and CE measure suitable for application in experiments to characterize entanglement. We further present the CE of some key higher-dimensional states such as qudit states and entangled optical states to validate the CE as a higher-dimensional measure of entanglement.

This article is part of the theme issue 'The quantum theory of light'.

# 1. Introduction

Entanglement is considered an essential resource in the field of quantum information [1]. Multipartite entanglement has unique uses such as quantum teleportation via multipartite entanglement channels [2] and network coding for secure key distribution [3]. Its importance has generated great interest in methods for practical detection of multipartite entanglement, of which the most common currently are entanglement witnesses and quantum state tomography [1,4]. The latter requires many measurements on a large ensemble of identical states, which scales exponentially with system size, and therefore, the search for more feasible schemes is ongoing. Entanglement witnesses for an $n$-qubit state require far fewer measurements but must be optimized for the state under consideration [5].

The *concentratable entanglement* (CE) [6], a multipartite measure of entanglement, can be directly estimated using the controlled SWAP test [7] or Bell basis measurements [8]. These methods can be applied to any $n$-qubit pure state so long as a source of (near) identical copies is available [7]. As well as estimating the concentratable entanglement of any subsystem, said tests can be used to identify which subsystems are entangled and what class of entanglement the entangled states belong to [6,7]. Since these tests are parallelized, errors and resource requirements scale linearly with system size, and therefore are particularly suited to multipartite states [7,8]. Further, multipartite measures of entanglement are often mathematically abstract, whereas CE has a clear operational meaning proportional to the number of 'concentrated' maximally entangled Bell pairs that can be extracted from two copies of the entangled state [6].

We build on this prior work by extending the analysis of mixed input states and exploring the validity of concentratable entanglement for higher-dimensional states. In an experimental setting, even states intended to be pure will be slightly mixed. Optical states, especially squeezed states, are important in the fields of quantum metrology [9], imaging [10] and computing [11], and qudits have the potential for increased quantum computing power and fault tolerance [12]. Overall, our work generalizes concentratable entanglement towards its use in practical applications.

The paper proceeds as follows. Section 2 provides a background on the states considered, namely qubit, qudit and coherent states. Next, we describe the entanglement monotone concentratable entanglement (CE) as defined in [6] and [8], and the experimentally implementable tests for estimating the CE of an ensemble of qubit states as described in [7] and [8]. In Section 3, the analytically calculated outputs of these tests on non-identical ensembles of mixed qubit states are investigated, leading to new definitions of the bounds on experimentally estimated concentratable entanglement. Finally, in Section 4, we discuss both the appropriate experimental methods for estimating the CE of higher-dimensional states and the validity of said CEs in these contexts. We summarize and conclude in Section 5.

# 2. Background

## (a) Qubits and entanglement

A qubit is analogous to a classical binary bit, but can also be in a superposition of the computational basis states $|0\rangle$ and $|1\rangle$. If a pure state, one qubit can be represented by the quantum state vector [13]

$$\begin{aligned}|\psi_1\rangle &= A_0|0\rangle + A_1|1\rangle \\ &= \begin{bmatrix} A_0 \\ A_1 \end{bmatrix},\end{aligned} \tag{2.1}$$

where $A_0, A_1 \in \mathbb{C}$, where the probability of measuring state $|k\rangle$ is $P(|k\rangle) = |A_k|^2$. The general two-qubit pure state has the state vector [13]

$$|\psi_2\rangle = A_{00}|00\rangle + A_{01}|01\rangle + A_{10}|10\rangle + A_{11}|11\rangle, \tag{2.2}$$

where $|j\rangle \otimes |k\rangle \equiv |jk\rangle$.

In the quantum circuit model of computation, reversible state transformations are represented as quantum logic gates. An operator $A$ representing a gate [13] acts on a vector $|\psi\rangle$ with $A|\psi\rangle = |\psi'\rangle$. Gates relevant to this work include the single-qubit Hadamard gate H [13]:

$$\mathrm{H} = \frac{1}{\sqrt{2}}\begin{bmatrix} 1 & 1 \\ 1 & -1 \end{bmatrix}, \tag{2.3}$$

which operates on a computational basis states such that

$$\mathrm{H}|0\rangle = \frac{1}{\sqrt{2}}(|0\rangle + |1\rangle) \quad \text{and} \quad \mathrm{H}|1\rangle = \frac{1}{\sqrt{2}}(|0\rangle - |1\rangle).$$

The two-qubit CNOT gate [13] flips the target qubit if the control qubit is in state $|1\rangle$. It has a matrix

$$\mathrm{CNOT} = \begin{bmatrix} 1 & 0 & 0 & 0 \\ 0 & 1 & 0 & 0 \\ 0 & 0 & 0 & 1 \\ 0 & 0 & 1 & 0 \end{bmatrix}, \tag{2.4}$$

where the first qubit is the control and the second is the target. The three-qubit Toffoli gate flips the target qubit if and only if the two control qubits are both in state $|1\rangle$ [13]. It can be represented by the matrix

$$\mathrm{T} = \begin{bmatrix} 1 & 0 & 0 & 0 & 0 & 0 & 0 & 0 \\ 0 & 1 & 0 & 0 & 0 & 0 & 0 & 0 \\ 0 & 0 & 1 & 0 & 0 & 0 & 0 & 0 \\ 0 & 0 & 0 & 1 & 0 & 0 & 0 & 0 \\ 0 & 0 & 0 & 0 & 1 & 0 & 0 & 0 \\ 0 & 0 & 0 & 0 & 0 & 1 & 0 & 0 \\ 0 & 0 & 0 & 0 & 0 & 0 & 0 & 1 \\ 0 & 0 & 0 & 0 & 0 & 0 & 1 & 0 \end{bmatrix}, \tag{2.5}$$

where the first two qubits are the controls and the third qubit is the target.

A multiple qubit system that cannot be expressed as a tensor product of its composite states is said to be entangled. The class of maximally bipartite entangled states are known as Bell states [13]:

$$|\Phi^{\pm}\rangle = \frac{|00\rangle \pm |11\rangle}{\sqrt{2}}, \quad |\Psi^{\pm}\rangle = \frac{|01\rangle \pm |10\rangle}{\sqrt{2}}. \tag{2.6}$$

These states are equivalent under local operations and classical communication (LOCC) and so we refer to their class set with $|\mathrm{Bell}\rangle \in \{|\Phi^{\pm}\rangle, |\Psi^{\pm}\rangle\}$. For example, $\mathrm{NOT}|\Phi^{+}\rangle = |\Psi^{+}\rangle$, where $\mathrm{NOT}|0\rangle = |1\rangle$ and $\mathrm{NOT}|1\rangle = |0\rangle$. Any valid measure of entanglement will have the same maximal value for any Bell state.

The Bell states form a two-qubit basis, the Bell basis, related to the computational basis by

$$(\mathrm{H} \otimes \mathrm{I})\mathrm{CNOT}\{|\Phi^{+}\rangle, |\Psi^{+}\rangle, |\Phi^{-}\rangle, |\Psi^{-}\rangle\} = \{|00\rangle, |01\rangle, |10\rangle, |11\rangle\}. \tag{2.7}$$

Therefore, if a state cannot be measured in the Bell basis directly, which is common for experiments, applying a Hadamard gate, a CNOT gate, and measuring each qubit in the computational basis will perform the Bell measurement.

For states with a greater number of qubits $n$, the classification of entangled states is richer than in the bipartite case, and multiple distinct classes of entanglement exist [14]. One class of multipartite entangled states are GHZ states, for example, [15]

$$|\mathrm{GHZ}_n\rangle = \frac{1}{\sqrt{2}}(|0\rangle^n + |1\rangle^n), \tag{2.8}$$

where $|0\rangle^n$ indicates $n > 2$ qubits all in state $|0\rangle$. If any qubits in a GHZ state are measured, all entanglement is destroyed. The unique class of $n$-qubit maximally entangled GHZ states [16] can be obtained by applying any reversible LOCC gates to $|\mathrm{GHZ}_n\rangle$. GHZ-like states are states with GHZ-type entanglement but are less entangled than $|\mathrm{GHZ}_n\rangle$, for example $|\psi\rangle = \sin\,\theta|0\rangle^n + \cos\,\theta|1\rangle^n$, where $\theta \neq \frac{\pi}{4}$.

A second unique multipartite class are W states, for example, a maximally entangled $n$-qubit W state [17]:

$$|\mathrm{W}_n\rangle = \frac{1}{\sqrt{n}}\sum_{k=1}^{n}|0\ldots1_k\ldots0\rangle_n, \tag{2.9}$$

where the subscript of $|1\rangle$ indicates its position in the $n$-qubit state $|x_n\ldots x_3x_2x_1\rangle_n$ such that for example $|0\ldots1_2\ldots0\rangle_n = |0010\rangle$. By some entanglement measures the maximally entangled W states are less entangled than maximally entangled GHZ states, 'maximal' meaning only within its class. However, W-like states are more robust as a measurement of one qubit does not destroy the entanglement of the rest of the state.

In practice, states are generally not pure, either because they are part of a larger state, or due to decoherence. These mixed states cannot be represented as a single ket vector as above but take the form of a density matrix [13]:

$$\rho = \sum^{T} p_i|\psi_i\rangle\langle\psi_i|, \tag{2.10}$$

where each element in the set $\{|\psi_1\rangle, |\psi_2\rangle, \ldots, |\psi_T\rangle\}$ is a unique pure $n$-qubit state and $p_i$ is the probability of $|\psi_i\rangle$ in the ensemble $\rho$ (therefore $\sum_i^K p_i = 1$). If $T = 1$, $\rho$ is a pure state, otherwise it is a mixed state. Gate $A$ acts on a density matrix $\rho$ with $A\rho A^\dagger = \rho'$ [13].

The *purity* of $\rho$ is given by [18]

$$\gamma = \mathrm{Tr}\rho^2, \tag{2.11}$$

where pure states have a purity of 1. A maximally mixed state of the form $\rho = \frac{I_N}{N}$, where $I_N$ is an identity matrix of size $N = D^n$ where $D = 2$, has purity $\gamma = \frac{1}{N}$. The purity of a state characterizes the available information about the quantum system [19]. Werner states [20] are of the form

$$\rho = (1-p)|\Psi\rangle\langle\Psi| + p\frac{I_N}{N} \tag{2.12}$$

and therefore the purity of $\rho$ is dependent on $p$, where $\rho$ is pure when $p = 0$ and maximally mixed when $p = 1$. Further, $\rho$ is separable [21] when $1 - p \leq (1 + D^{n-1})^{-1}$. The reduced density matrix of state $R$ within composite system $RT$ is given by the partial trace [13]

$$\rho_R = \mathrm{Tr}_T(\rho_{RT}) = \sum_k {}_T\langle k|\rho_{RT}|k\rangle_T\,. \tag{2.13}$$

Entanglement in a bipartite pure state is now well understood as the degree of mixedness of each subsystem, where the 'mixedness' characterizes a lack of information about the state of a quantum system [19]. Any entanglement measure between subsystem $A$ and subsystem $B$ therefore should increase as the purity of $\rho_A$ and $\rho_B$ decrease.

For example, the entanglement in a pure state can be quantified by the *entropy of entanglement*, given by the von Neumann entropy $S_V$ of the reduced density matrix representing each subsystem, such that [13]

$$S_V(\rho_A) = -\mathrm{Tr}\,[\rho_A \log\rho_A], \tag{2.14}$$

where $\mathrm{Tr}\,[A]$ is the trace of a matrix $A$. For two-qubit states, the concurrence $C_2$ quantifies entanglement and is given by [22,23]

$$C_2(|\psi\rangle_{AB}) = \sqrt{2(1-\mathrm{Tr}\rho_A^2)} \tag{2.15}$$

$$C_2(\rho_{AB}) \geq \sqrt{2\,\mathrm{tr}\left[\rho^2\right] - \mathrm{tr}\left[\rho_A^2\right] - \mathrm{tr}\left[\rho_B^2\right]} \tag{2.16}$$

with $0 \leq C_2 \leq 1$. A concurrence of $C_2 = 0$ indicates that no entanglement is present in the system and $C_2 = 1$ corresponds to a maximally entangled state [22]. The experimental estimation of concurrence (and therefore the bipartite equivalent of the techniques described in this article) has been well studied [24–26].

## (b) Higher-dimensional states

### (i) Qudits

Qudits behave similarly to qubits, but are of a higher dimension and are therefore not restricted to superpositions of the 0 and 1 binary states. A general one-qudit pure state is of the form

$$|\psi_{D,n=1}\rangle = \sum_{k=0}^{D-1} A_k|k\rangle, \tag{2.17}$$

where $D > 2$ and $D \in \mathbb{Z}^+$ is the dimension of the qudit. The $D = 3$ case is known as a *qutrit* [27,28]. Higher dimensions allow the possibility for richer quantum architecture and simulation [29], simplified quantum circuits [30] and higher fault tolerance [31]. Entanglement in qudits is defined similarly to the qubit case [27]; for example, $\left|\Phi^+_{D=3,n=2}\right\rangle = \frac{1}{\sqrt{3}}(|00\rangle + |11\rangle + |22\rangle)$ is a maximally entangled two-qutrit state.

### (ii) Optical states

One of the most important optical states is the coherent state $|\alpha\rangle$, which is the unique eigenstate of the annihilation operator $\hat{a}$ in a quantum harmonic oscillator [32]:

$$\hat{a}|\alpha\rangle = \alpha|\alpha\rangle, \tag{2.18}$$

where $\alpha$ is a complex amplitude $\alpha = |\alpha|e^{i\phi}$. Coherent states follow a Poisson number distribution when represented based on photon number states [33], or Fock states, $|n\rangle$:

$$|\alpha\rangle = e^{-\frac{|\alpha|^2}{2}} \sum_{n=0}^{\infty} \frac{\alpha^n}{\sqrt{n!}}|n\rangle, \tag{2.19}$$

where $|\alpha|^2 = \mu$ is the average number of photons. It follows that the probability of finding $m$ photons is $P(m) = |\langle m|\alpha\rangle|^2 = \mu^m e^{-\mu}/m!$.

A coherent state can also be thought of as the vacuum state $|0\rangle$ displaced to a location $\alpha$ in phase space, due to the action of a displacement operator $\hat{D}(\alpha)$ such that [32]

$$|\alpha\rangle = e^{\alpha\hat{a}^\dagger - \alpha^*\hat{a}}|0\rangle = \hat{D}(\alpha)|0\rangle. \tag{2.20}$$

In contrast to the photon number states, coherent states are not orthogonal and form an overcomplete basis. The inner product between coherent states $|\alpha\rangle$ and $|\beta\rangle$ is given by [33]

$$\langle\alpha|\beta\rangle = e^{-\frac{1}{2}|\alpha|^2 - \frac{1}{2}|\beta|^2 + \alpha^*\beta}. \tag{2.21}$$

The quadrature operators $\hat{X} = \frac{1}{2}(\hat{a}^\dagger + \hat{a})$ and $\hat{Y} = \frac{i}{2}(\hat{a}^\dagger - \hat{a})$ obey the uncertainty relation [32]

$$\langle(\Delta\hat{X})^2\rangle\langle(\Delta\hat{Y})^2\rangle \geq \frac{1}{16},$$

where $\langle(\Delta O)^2\rangle \equiv \langle\hat{O}^2\rangle - \langle\hat{O}\rangle^2$ is the variance of $\hat{O}$ and $\langle\hat{O}\rangle = \langle\psi|\hat{O}|\psi\rangle$. Coherent states minimize this with $\langle(\Delta\hat{X})^2\rangle\langle(\Delta\hat{Y})^2\rangle = \frac{1}{16}$ and are therefore minimum uncertainty states. A state is said to be squeezed whenever $\langle(\Delta\hat{X})^2\rangle < \frac{1}{4}$ or $\langle(\Delta\hat{Y})^2\rangle < \frac{1}{4}$, with one quadrature's uncertainty 'squeezed' at the expense of the other [32].

Squeezing can be performed over multiple modes [32]. The two-mode squeezing operator applied to modes $\hat{a}$ and $\hat{b}$ is

$$\hat{S}_2(\xi) = \exp(\xi^*\hat{a}\hat{b} - \xi\hat{a}^\dagger\hat{b}^\dagger), \tag{2.22}$$

for $\xi = re^{i\theta}$, where $r$ is known as the squeeze parameter and $\theta$ indicates the direction of squeezing [34]. Therefore, a general two-mode squeezed coherent state $|\alpha, \xi\rangle$ can be written as [32]

$$|\alpha, \beta, \xi\rangle = \hat{D}(\alpha)\hat{D}(\beta)\hat{S}_2(\xi)|00\rangle \tag{2.23}$$

using the displacement operator from equation (2.20). Since $\hat{S}_2(\xi)$ cannot be written as a product of two single-mode squeeze operators $\hat{S}_\alpha(\xi) = \exp[\frac{1}{2}(\xi^*\hat{a}^2 - \xi\hat{a}^{\dagger 2})]$ and $\hat{S}_\beta(\xi)$, this squeezing entangles the two modes. It can be shown that the entropy of entanglement increases [34,35] with the squeeze parameter $r$.

Cat states are linear superpositions of coherent states with phase differences. They are of particular interest due to their applicability in quantum computing [11] and as the building blocks for entangled coherent states [11,36]. Entangled coherent states (ECSs) exhibit entanglement between modes of the electromagnetic field. They have applications across a range of fields such as quantum optics [11], quantum information processing [36] and quantum metrology [9]. ECSs are also fundamentally interesting as entangled macroscopic states with minimized uncertainty. We will consider two-mode entangled coherent states of the form

$$|ECS_{\alpha,\beta}\rangle = \mathcal{N}_{\alpha,\beta}(A_{\alpha\alpha}|\alpha\rangle|\alpha\rangle + A_{\alpha\beta}|\alpha\rangle|\beta\rangle + A_{\beta\alpha}|\beta\rangle|\alpha\rangle + A_{\beta\beta}|\beta\rangle|\beta\rangle), \tag{2.24}$$

where

$$\begin{aligned}\frac{1}{\mathcal{N}^2_{\alpha,\beta}} &= |A_{\alpha\alpha}|^2 + |A_{\alpha\beta}|^2 + |A_{\beta\alpha}|^2 + |A_{\beta\beta}|^2 \\ &\quad + \langle\alpha|\beta\rangle A^*_{\alpha\alpha}(A_{\alpha\beta} + A_{\beta\alpha} + \langle\alpha|\beta\rangle A_{\beta\beta}) + \langle\alpha|\beta\rangle A^*_{\alpha\beta}(A_{\alpha\alpha} + \langle\alpha|\beta\rangle A_{\beta\alpha} + A_{\beta\beta}) \\ &\quad + \langle\alpha|\beta\rangle A^*_{\beta\alpha}(A_{\alpha\alpha} + \langle\alpha|\beta\rangle A_{\alpha\beta} + A_{\beta\beta}) + \langle\alpha|\beta\rangle A^*_{\beta\beta}(\langle\alpha|\beta\rangle A_{\alpha\alpha} + A_{\alpha\beta} + A_{\beta\alpha}).\end{aligned} \tag{2.25}$$

A useful example is the ECS $|ECS_{\alpha,-\alpha}\rangle$ where $\beta = -\alpha$ which can be implemented through parametric amplification and photodetection [36]. The greater the value of $\alpha$ the smaller the overlap $\langle\alpha|-\alpha\rangle = e^{-2|\alpha|^2}$, and therefore the more distinguishable the states $|\alpha\rangle$ and $|\beta\rangle$ are from one another.

## (c) Binomial distribution

Let there be a register of random variables from $M$ trials $\{X_1, ..., X_M\}$ such that $X \in \{x, x'\}$, that follow the binomial distribution $B(M, P(x))$, where $P(x)$ is the probability of $x$. The probability of getting exactly $k$ instances of $X = x$ in $M$ trials is given by the probability mass function [37]:

$$f(k, M, P(x)) = \binom{M}{k} P(x)^k (1 - P(x))^{M-k}. \tag{2.26}$$

According to the Maximum Likelihood Estimation [37] procedure, $\tilde{P}(x)$, the most likely value of $P(x)$, maximizes $f(k, M, P(x))$. This gives simply:

$$\tilde{P}(x) = \frac{k}{M}. \tag{2.27}$$

We define the mean [37] error of $\tilde{P}(x)$ in terms of the fractional error $\frac{|\tilde{P}(x) - P(x)|}{P(x)}$ as

$$Err(M, P(x)) = \sum_{k=0}^{M} f(k, M, P(x)) \frac{|\frac{k}{M} - P(x)|}{P(x)}. \tag{2.28}$$

### (d) Concentratable entanglement

Denote $S = \{1,2,\ldots,n\}$ as the set of labels for each qubit in input state $\rho = |\psi\rangle\langle\psi|$ and $\mathcal{P}(S)$ as its power set (the set of all subsets). For any set of qubit labels, $s \in \mathcal{P}(S)\setminus\{\varnothing\}$ (where $\{\varnothing\}$ is the empty set), the CE is [6]

$$\mathcal{C}_{|\psi\rangle}(s) = 1 - \frac{1}{2^{c(s)}}\sum_{\alpha\in\mathcal{P}(s)}\gamma_\alpha, \tag{2.29}$$

where $c(s)$ is the cardinality of the sets and $\gamma_\alpha = \mathrm{Tr}\rho_\alpha^2$ is the purity of the joint reduced state $\rho_\alpha = \mathrm{Tr}_{S\setminus\alpha}\rho$. In short, the lower each of the joint reduced purities $\{\gamma_\alpha : \alpha \in \mathcal{P}(s)\}$ the greater $\mathcal{C}_{|\psi\rangle}(s)$.

Local purities $\mathrm{tr}[\rho_\alpha^2]$ (in any possible partition) cannot decrease, on average, under local or separable operations [6]. Since it is entirely in terms of local purities, pure state concentratable entanglement $\mathcal{C}_{|\psi\rangle}(s)$ is also non-increasing, on average, under LOCC: a requirement for entanglement measures.

Concentratable entanglement can be estimated using the controlled SWAP (c-SWAP) test for entanglement [6,7]. During each round, the circuit in figure 1*a* is applied to the *k*th subsystems in the *n*-partite state $\rho$ and its (near) copy $\rho'$. The c-SWAP gate, with matrix

$$\text{c-SWAP} = \begin{bmatrix} 1&0&0&0&0&0&0&0\\ 0&1&0&0&0&0&0&0\\ 0&0&1&0&0&0&0&0\\ 0&0&0&1&0&0&0&0\\ 0&0&0&0&1&0&0&0\\ 0&0&0&0&0&0&1&0\\ 0&0&0&0&0&1&0&0\\ 0&0&0&0&0&0&0&1 \end{bmatrix}, \tag{2.30}$$

performs a swap between state $\rho_k$ and state $\rho'_k$ for each element in their superpositions, controlled on a qubit ancilla $C$. Note that even if $\rho = \rho' \to \rho_k = \rho'_k$, $\mathrm{SWAP}(\rho_k \otimes \rho'_k)\mathrm{SWAP}^\dagger \neq \rho_k \otimes \rho'_k$ unless $\rho_k$ is pure and therefore there is no entanglement between $\rho_k$ and the rest of $\rho$. The c-SWAP gate is not usually directly implementable in a physical setting but can be decomposed into CNOTs from equation (2.4) and Toffolis from equation (2.5), shown in figure 1*b* for $n = 2$. Then each of the $n$ ancillas is measured in the computational basis, giving joint output $\sigma_C = \{\sigma_{C_1}, \sigma_{C_2}, \ldots, \sigma_{C_n}\}$. Since this scheme is probabilistic, the circuit needs to be repeated $M$ times and therefore requires $2M$ (near) copies of $\rho$. After $M$ rounds, the probability of obtaining any given string $\in \{0,1\}^n$ from $\sigma_C$ can be calculated from the final data string $\{\sigma_C^{(m)} : m \in \{1,\ldots,M\}\}$.

Assuming an ensemble of identical states $\rho = \rho'$ for all $M$ rounds, the CE of pure state $|\psi\rangle$ is equivalent to the c-SWAP test results [6]

$$\begin{aligned}\mathcal{C}_{|\psi\rangle}(s) &= 1 - \sum_{z\in\mathcal{Z}_0(s)} P(z)\\ &= \sum_{z\in\mathcal{Z}_1^{\mathrm{even}}(s)} P(z)\\ &\equiv P(\mathcal{Z}_1^{\mathrm{even}}(s)),\end{aligned} \tag{2.31}$$

where $P(z)$ is the probability of measuring $z$ on $\sigma_C$, $\mathcal{Z}_0(s)$ is the set of all bitstrings with $|0\rangle$s on all indices in $s$ and $\mathcal{Z}_1^{\mathrm{even}}(s)$ is the set of bit strings with even Hamming weight and with at least one $|1\rangle$ on an index in $s$. Concurrence [22], generalized concurrence [38], the $n$-tangle [39,40], and linear entropy of entanglement [13] are all special cases of concentratable entanglement [6,41].

The c-SWAP test for entanglement is a state comparison test [42–44] performed on each subsystem's reduced state: the more entangled the overall state, the less pure each subsystem's reduced state, the greater the effect of the SWAP gate on those subsystems, and therefore, the greater the probability of measuring $|1\rangle_C$ on the corresponding control qubits. Specifically, a measurement of $|1\rangle_{C_j} \otimes |1\rangle_{C_k}$ evidences low purity of both $\rho_j$ and $\rho_k$, and therefore entanglement between subsystems $\rho_j$ and $\rho_k$. Alternatively, we can consider the test's effect on the test states: if the final states of channels $A$ and $B$ were to be measured, we would find that the probability of measuring $|1\rangle_{C_j} \otimes |1\rangle_{C_k}$ is equivalent to the probability of measuring $|\Psi^-\rangle_{A_jB_j} \otimes |\Psi^-\rangle_{A_kB_k}$, where $|\Psi^-\rangle$ is the 'singlet' Bell state from equation (2.6). Therefore, the c-SWAP test 'concentrates' entanglement between $\rho_j$ and $\rho_k$ into singlet states between $\rho_j$ &$\rho'_j$ and between $\rho_k$ &$\rho'_k$, then flags these concurrent singlet states with $|1\rangle_{C_j} \otimes |1\rangle_{C_k}$. The total entanglement $\mathcal{C}_{|\psi\rangle}(s)$ is then the linear combination of the probabilities of each instance of an even number of $|1\rangle_C$s, $P(\mathcal{Z}_1^{\text{even}}(s))$. For a more in-depth discussion of the relationship between CE and the c-SWAP probability results, see Foulds [41].

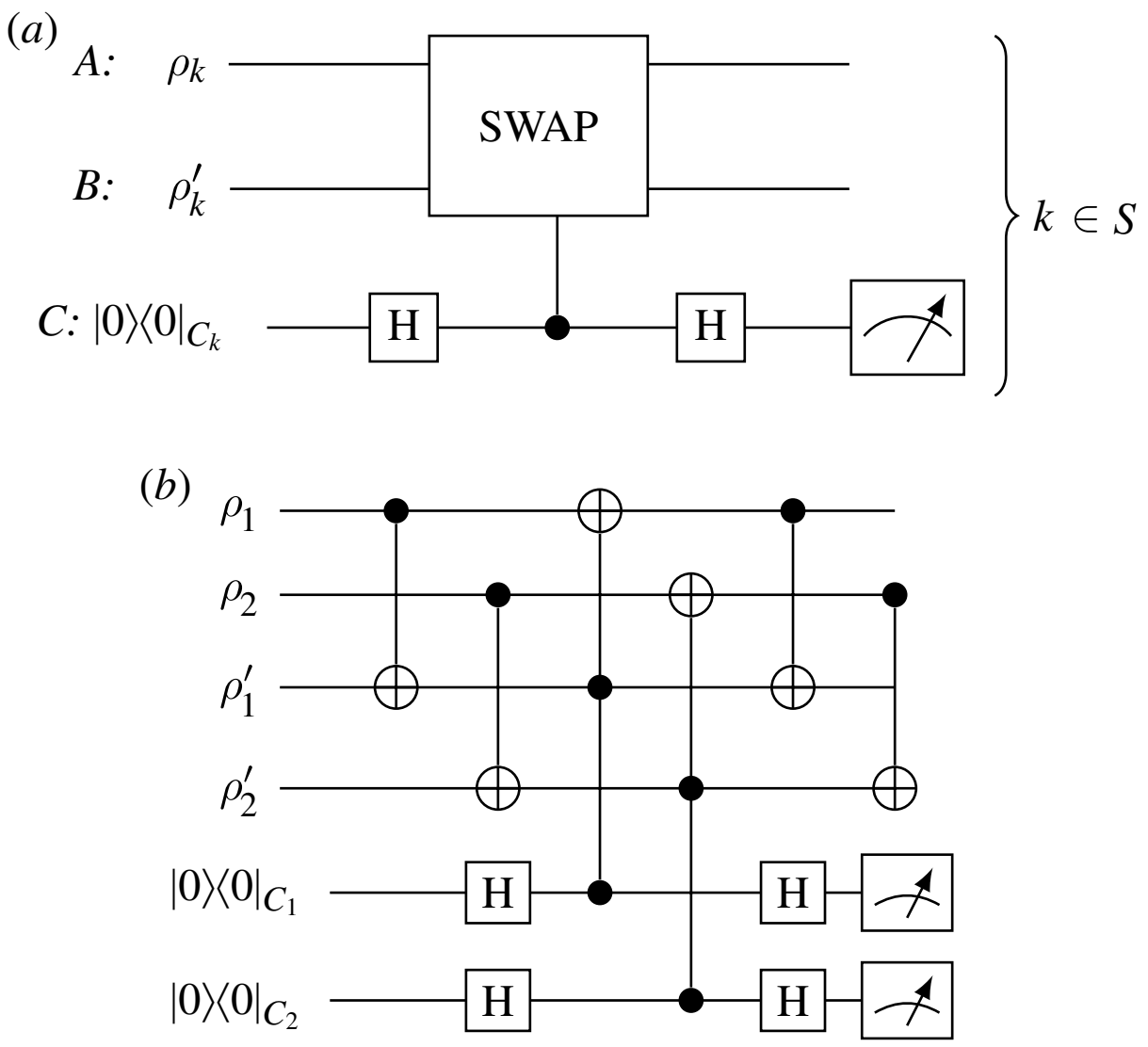

**Figure 1.** *(a)* shows one round of the c-SWAP test for entanglement on the input states $\rho$ and $\rho'$. The circuit can be considered in parallel for each qubit, with the SWAP gate applied to the $k$th qubit in each of $\rho$ and $\rho'$ controlled on the $k$th control qubit for $k \in S = \{1, 2, \ldots, n\}$. *(b)* shows the gate breakdown for the two-qubit version of the test. The two-qubit gates are CNOT gates and the three-qubit gates are Toffolis. The second set of CNOTs are optional, they do not affect the measurement outcomes.

However, in the experiment, the elements of the input state ensemble will not be identical. If $\rho = |\psi\rangle\langle\psi|$ and $\rho' = |\phi\rangle\langle\phi|$ where $|\psi\rangle \neq |\phi\rangle$, there will be a non-zero probability of measuring an odd number of $|1\rangle$s in the control ancillas. Therefore [7]

$$\begin{aligned}\mathcal{C}_{|\psi\rangle,|\phi\rangle}(s) &= 1 - \sum_{z \in \mathcal{Z}_0(s)} P(z)\\ &\equiv P(\mathcal{Z}_1^{\text{even}}(s)) + P(\mathcal{Z}_1^{\text{odd}}(s)).\end{aligned} \tag{2.32}$$

Equivalently, the concentratable entanglement of qubit states can be obtained via Bell basis measurements [8] that do not require three-qubit gates or ancilla qubits. As with the c-SWAP test the Bell basis measurement test requires $2M$ near copies of the $n$-partite input state $\rho$. For each round $m \in \{1, \ldots, M\}$, the circuit in figure 2 is applied to the $k$th subsystems in $\rho$ and its pair $\rho'$. All $2n$ qubits are measured, with output $\sigma_{AB} = \{\sigma_{A_1}, \sigma_{B_1}, \sigma_{A_2}, \sigma_{B_2}, \ldots, \sigma_{A_n}, \sigma_{B_n}\} \equiv \{\sigma_{AB_1}, \sigma_{AB_2}, \ldots, \sigma_{AB_n}\}$. After $M$ rounds of the test the output data $\sigma_{AB}^{(m)}$ are obtained. The probability results for the Bell measurement test relate exactly to the probability results for the c-SWAP test [8] with

$$P_{\text{c-SWAP}}(|1\rangle_C) \equiv P_{\text{Bell basis}}(|11\rangle_{AB}), \tag{2.33}$$

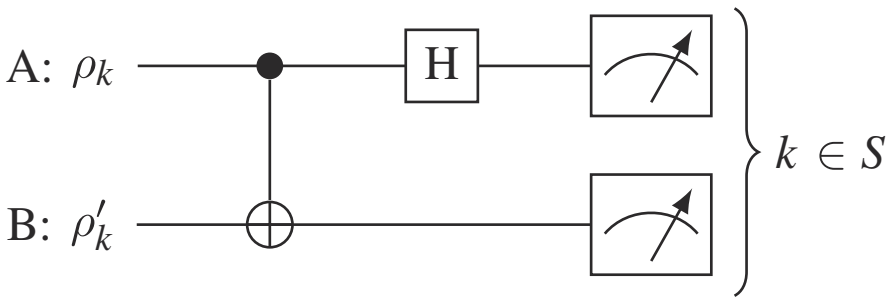

**Figure 2.** One round $m$ of the Bell basis test for entanglement on input states $\rho$ and $\rho'$. H is a Hadamard gate and the two-qubit gate is a CNOT gate.

$$P_{\text{c-SWAP}}(|0\rangle_C) \equiv P_{\text{Bell basis}}(|00\rangle_{AB} \cup |01\rangle_{AB} \cup |10\rangle_{AB}),$$

where $P_{\text{c-SWAP}}(z_C)$ is the probability of measuring the control in state $z$ after the c-SWAP test is applied to $\rho$, and $P_{\text{Bell basis}}(zz'_{AB})$ is the probability of measuring the final states in state $zz'$ after the c-SWAP test is applied to $\rho$. Therefore, the CE of $|\psi\rangle$ is equivalent to

$$\begin{aligned}\mathcal{C}_{|\psi\rangle}(s) &= 1 - P(\mathcal{Z}_0(s)) \\ &= P(\mathcal{Z}_1^{\text{even}}(s))\end{aligned},$$

where $P(z)$ is the probability of measuring $z$ on $\sigma_{AB}$, $\mathcal{Z}_0(s)$ is the set of all bit strings with $|00\rangle \cup |01\rangle \cup |10\rangle$ on all indices in $s$ and $\mathcal{Z}_1^{\text{even}}(s)$ is the set of bit strings with an even, non-zero number of $|11\rangle$s on an index in $s$. Therefore, the CE is equivalently recovered by the Bell basis measurement test for qubit input states. For a discussion on the size of $M$ relative to the precision of $\mathcal{C}_{|\psi\rangle}$, see Beckey *et al.* [8].

This method further illuminates concentratable entanglement's operational meaning. During the Bell basis measurement test, the subsystems $\rho_k$ and $\rho'_k$ undergo the operation to convert Bell states into the computational basis states. Therefore, a measurement of $|11\rangle_{AB}$ signifies the 'singlet' Bell state $|\Psi^-\rangle$ between input states $\rho_k$ and $\rho'_k$ [8,41].

The Bell basis measurement method, numerically simulated with noisy Rydberg-mediated gates, was shown to be less lossy and therefore more efficient than the c-SWAP test for the estimation of the CE of a pure $n$-qubit GHZ state [8].

Foulds *et al.* [7] and Beckey *et al.* [8] discuss the cases where the input states are non-identical and the input states are mixed respectively. In either case $P(\mathcal{Z}_1^{\text{odd}}(s)) \neq 0$, increasing with decreased input states fidelity and/or purity, and therefore, the CE must be redefined to accommodate this experimental reality. Beckey *et al.* [8] combine the mixed state extension for the two-qubit concurrence with the fact that the pure state CE can be written as a sum over all pairs in $S$ to define the lower bound of the CE of a mixed state. In the case of total CE (where $s = S$), this is

$$\mathcal{C}_\rho^l(S) = \frac{1}{2^n} + \left(1 - \frac{1}{2^n}\right)\text{tr}\left[\rho^2\right] - \frac{1}{2^n}\sum_{\alpha \in \mathcal{P}(S)} \text{tr}\left[\rho_\alpha^2\right]. \tag{2.34}$$

The superscript $l$ denotes a lower bound.

## 3. Concentratable entanglement for qubit mixed states extended

In previous work, non-identical pure input states [7] and the general case of identical mixed input states have been considered [6], but not specific examples of identical mixed input states or non-identical mixed input states (undoubtedly the experimental reality). Here, we present analytical results of the c-SWAP test for GHZ-like and W-like Werner states and investigate their corresponding CE values to extend CE to experimental use. For the code used to generate the results see github.com/sfoulds [45].

The pure state CE is defined as

$$\mathcal{C}_{|\psi\rangle}(s) = 1 - \frac{1}{2^{c(s)}} \sum_{\alpha \in \mathcal{P}(s)} \mathrm{tr}\left[\rho_\alpha^2\right]. \tag{3.1}$$

When the input states to the above entanglement tests are mixed, i.e. $\rho \neq |\psi\rangle\langle\psi|$, this value corresponds to

$$\begin{aligned} 1 - \frac{1}{2^{c(s)}} \sum_{\alpha \in \mathcal{P}(s)} \mathrm{tr}\left[\rho_\alpha^2\right] &= 1 - P(\mathcal{Z}_0(s)) \\ &\equiv P(\mathcal{Z}_1^{\mathrm{even}}(s)) + P(\mathcal{Z}_1^{\mathrm{odd}}(s)), \end{aligned} \tag{3.2}$$

where $P(z)$ are the probability results calculated from outputs $\sigma^{(m)}$ of the c-SWAP test or Bell basis measurements as detailed in §2d on input states $\rho = \rho'$. We have calculated that these probabilities in turn are equivalent to

$$\begin{aligned} P(\mathcal{Z}_1^{\mathrm{even}}) &= \frac{1}{2}\left(1 + \mathrm{tr}\left[\rho^2\right]\right) - \frac{1}{2^n} \sum_{\alpha \in \mathcal{P}(S)} \mathrm{tr}\left[\rho_\alpha^2\right], \\ &\equiv \frac{1}{2}(1 + \gamma) - \frac{1}{2^n} \sum_{\alpha \in \mathcal{P}(S)} \gamma_\alpha, \end{aligned} \tag{3.3}$$

$$\begin{aligned} P(\mathcal{Z}_1^{\mathrm{odd}}) &= \frac{1}{2}\left(1 - \mathrm{tr}\left[\rho^2\right]\right) \\ &\equiv \frac{1}{2}(1 - \gamma) \end{aligned} \tag{3.4}$$

when $s = S$, which is the case that we will be covering.

Since mixed states are statistical mixtures—see equation (2.10)—their degree of entanglement must also be probabilistic. The minimum and maximum average of a mixed state's CE is therefore respectively [46]

$$\mathcal{C}_\rho^{\cup} = \inf \sum_i p_i \mathcal{C}_{|\psi_i\rangle}, \tag{3.5}$$

$$\mathcal{C}_\rho^{\cap} = \sup \sum_i p_i \mathcal{C}_{|\psi_i\rangle}, \tag{3.6}$$

where the infimum and supremum are over the possible decompositions of $\rho = \sum_i p_i |\psi_i\rangle\langle\psi_i|$. When $\rho$ is pure, these values converge. The optimizations required to solve these equations are hard to compute and so we look for alternative methods to find the upper and lower bounds $\mathcal{C}_\rho^l$ and $\mathcal{C}_\rho^u$ respectively, such that the state $\rho$ must have at least (most) $\mathcal{C}_\rho^l(\mathcal{C}_\rho^u)$ concentratable entanglement. The greater the purity of the state, the closer these bounds should be to one another.

Beckey *et al.* [8] find the lower bound of the total CE, equation (2.34), by extending the previously known lower bound of concurrence: equation (2.16). We have calculated that this lower bound in terms of entanglement test outcomes is

$$\mathcal{C}_\rho^l(S) = P(\mathcal{Z}_1^{\mathrm{even}}) - \left(1 - \frac{2}{2^n}\right) P(\mathcal{Z}_1^{\mathrm{odd}}). \tag{3.7}$$

We suppose that the upper bound of the CE of mixed states would similarly be in terms of $P(\mathcal{Z}_1^{\mathrm{even}})$ and $P(\mathcal{Z}_1^{\mathrm{odd}})$. Note that the value of $P(\mathcal{Z}_1^{\mathrm{even}})$ not only increases as the sum of the joint reduced purities decreases (as does $\mathcal{C}_{|\psi\rangle}$) but also increases as the purity of $\rho$ increases.

In this section, we investigate the analytical entanglement test results for mixed and non-identical input states in order to conjecture an upper bound for CE.

## (a) Identical input states

We model mixedness with Werner states of the form in equation (2.12). First we consider that the pair of mixed input states are identical to one another such that $\rho = \rho' = \rho_\Psi(p) = (1-p)|\Psi\rangle\langle\Psi| + p\frac{I_N}{N}$, where $0 \le p \le 1$: therefore the variable $p$ controls the mixedness of $\rho$ and its pair $\rho'$.

Let $\rho = \rho' = \rho_{\mathrm{GHZ}}(p) = (1-p)|\mathrm{GHZ}\rangle\langle\mathrm{GHZ}| + p\frac{I_N}{N}$, where $|\mathrm{GHZ}\rangle = A_0|0\rangle^n + A_1|1\rangle^n$ is a GHZ-like state and $|A_0|^2 + |A_1|^2 = 1$. If $A_0 = A_1 = \frac{1}{\sqrt{2}}$ then this is the 'maximally entangled' GHZ state from equation (2.8). The purity of $\rho_{\mathrm{GHZ}}$ is then $\gamma = 1 - \frac{2^n-1}{2^n}p(2-p)$. This state results in

$$P(\mathcal{Z}_1^{\mathrm{odd}}) = \frac{2^n-1}{2^{n+1}}p(2-p) \tag{3.8}$$

$$P(\mathcal{Z}_1^{\mathrm{even}}) = 4\left(\frac{1}{2} - \frac{1}{2^n}\right)A_0^2A_1^2 - \left[4\left(\frac{1}{2} - \frac{1}{2^n}\right)A_0^2A_1^2 - \frac{1}{2} - \frac{1}{2^{n+1}} + \left(\frac{3}{4}\right)^n\right]p(2-p) \tag{3.9}$$

and therefore

$$1 - P(\mathcal{Z}_0) = 4\left(\frac{1}{2} - \frac{1}{2^n}\right)A_0^2A_1^2 + \left[1 - 4\left(\frac{1}{2} - \frac{1}{2^n}\right)A_0^2A_1^2 - \left(\frac{3}{4}\right)^n\right]p(2-p) \tag{3.10}$$

$$\mathcal{C}^l_{\rho_{\mathrm{GHZ}}}(S) = 4\left(\frac{1}{2} - \frac{1}{2^n}\right)A_0^2A_1^2 - \left[4\left(\frac{1}{2} - \frac{1}{2^n}\right)A_0^2A_1^2 - \frac{2}{2^n} + \frac{1}{4^n} + \left(\frac{3}{4}\right)^n\right]p(2-p), \tag{3.11}$$

where $\mathcal{Z}^0 = |0\rangle_C^n$, shown in figure 3 for $A_0 = \frac{1}{\sqrt{2}}$. $P(\mathcal{Z}_1^{\mathrm{even}})$, $1 - P(\mathcal{Z}_0)$ and $\mathcal{C}^l_{\rho_{\mathrm{GHZ}}}(S)$ converge when $\gamma = 1$. For further examples see Foulds [41]: it can be seen that for a mixed W-like state the above equations can similarly be written in terms of $p(2-p) = (1-\gamma)\frac{2^n}{2^n-1}$.

For any (qubit) input state $P(\mathcal{Z}_1^{\mathrm{odd}}) = \frac{1}{2}(1-\gamma)$, and therefore the purity of $\rho$ can be directly estimated. Further, $P(\mathcal{Z}_1^{\mathrm{even}})$ and $\mathcal{C}^l_\rho(S)$ increase linearly with $\gamma$ and $1 - P(\mathcal{Z}_0)$ decreases linearly with $\gamma$. An ideal entanglement measure would correlate with the level of useful entanglement associated with $\rho$, and so we would expect said measure to *decrease* with decreased purity. This is true for $\mathcal{C}^l_\rho$ and $P(\mathcal{Z}_1^{\mathrm{even}})$. We would further expect an entanglement measure to give zero for a separable state. The separability criterion [21] for qubit Werner states is $\gamma \le \frac{2^n+8}{(2^n+2)^2} = \gamma_{\mathrm{separable}}$ and is shown in figure 3 with vertical lines. The test results when $\gamma = \gamma_{\mathrm{separable}}$ are shown in figure 4 for various $n$: the experimentally estimable values that are closest to zero are $\mathcal{C}^l_\rho(S)$ below and $P(\mathcal{Z}_1^{\mathrm{even}})$ above. Therefore $\mathcal{C}^l_\rho$ underestimates the entanglement in $\rho$. $P(\mathcal{Z}_1^{\mathrm{even}}) > 0$ for all $\gamma$ and so overestimates the entanglement in $\rho$ for low $\gamma$. This suggests $P(\mathcal{Z}_1^{\mathrm{even}})$ could be a tighter upper bound than $1 - P(\mathcal{Z}_0)$.

Shown in figure 5 is the mean error, from equation (2.28), of $\tilde{P}(\mathcal{Z}_1^{\mathrm{even}})$ calculated from $M$ trials of the Bell basis test for entanglement on $2M$ copies of $\rho_{\mathrm{Bell}(\gamma)}$. The error decreases with increased purity with $Err(M, P(\mathcal{Z}_1^{\mathrm{even}})) \approx 3^{-\log_{10}(M)}(1.1 - 2\log_{10}(\gamma))$. Since $P(\mathcal{Z}_1^{\mathrm{even}})$ is linear in $\gamma$ regardless of entanglement class, we conjecture that the mean error is logarithmic regardless of class.

## (b) Non-identical input states

Now consider non-identical mixed states, usually the experimental reality. In this work, we assume the two input states $\rho$ and $\rho'$ are similar enough such that their density matrices commute for mathematical simplicity.

Let us define a new value

$$L = P(\mathcal{Z}_1^{\mathrm{even}}) - \left(1 - \frac{2}{2^n}\right)P(\mathcal{Z}_1^{\mathrm{odd}}), \tag{3.12}$$

which is equal to $\mathcal{C}^l_\rho$ expressed in entangled test probabilities from equation (3.7). Since we are now considering non-identical input states, this equation is no longer equivalent to $\mathcal{C}^l_\rho$ from equation (2.34), expressed in terms of $\rho$. We will investigate how close the value $L$ obtained by the entanglement test is to $\mathcal{C}^l_\rho$ and $\mathcal{C}^l_{\rho'}$.

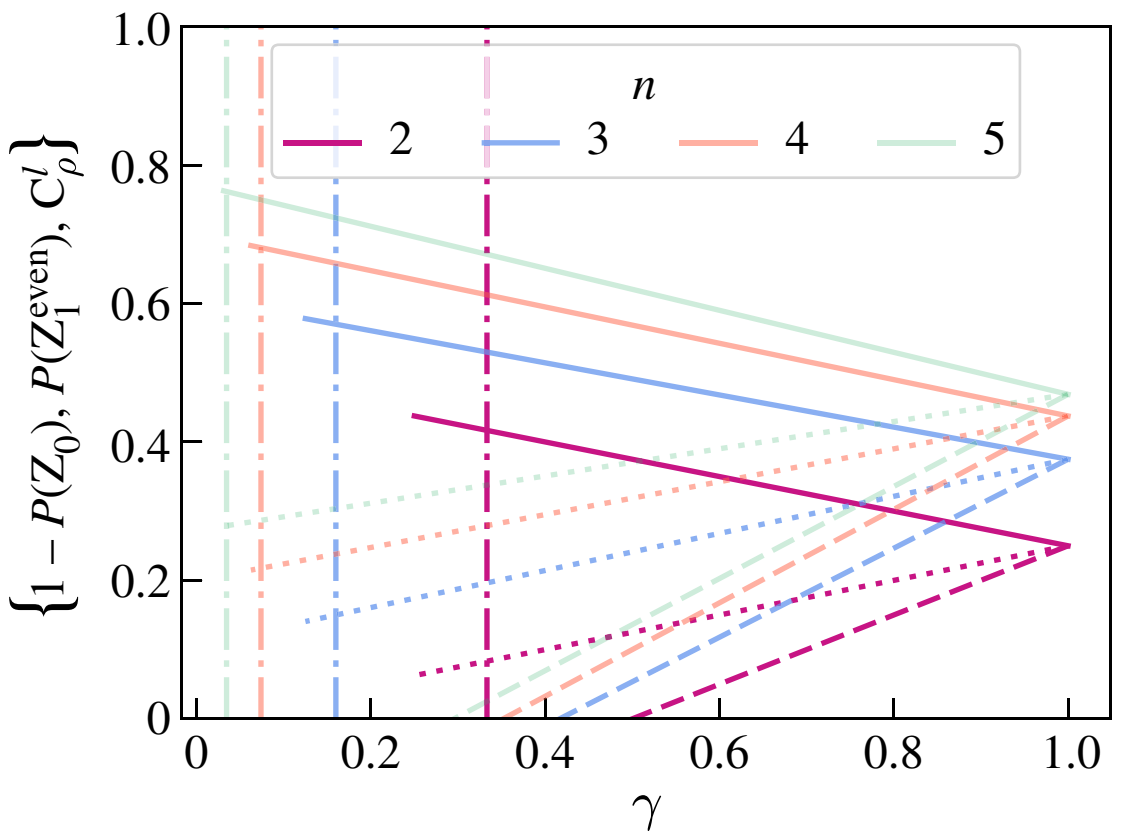

**Figure 3.** Test results for input states $\rho = \rho' = \rho_{\text{GHZ}}(p) = (1-p)|\text{GHZ}\rangle\langle\text{GHZ}| + p\frac{I_N}{N}$. Unbroken lines represent $1 - P(\mathcal{Z}_0)$, dotted lines $P(\mathcal{Z}_1^{\text{even}})$ and dashed lines $\mathcal{C}^l_{\rho_{\text{GHZ}}}$; $n = 2$ for opaque lines, with decreasing opacity for increasing $n$. Dash-dot lines show the value of $\gamma_{\text{separable}}$ such that the state is separable for $\gamma \le \gamma_{\text{separable}}$.

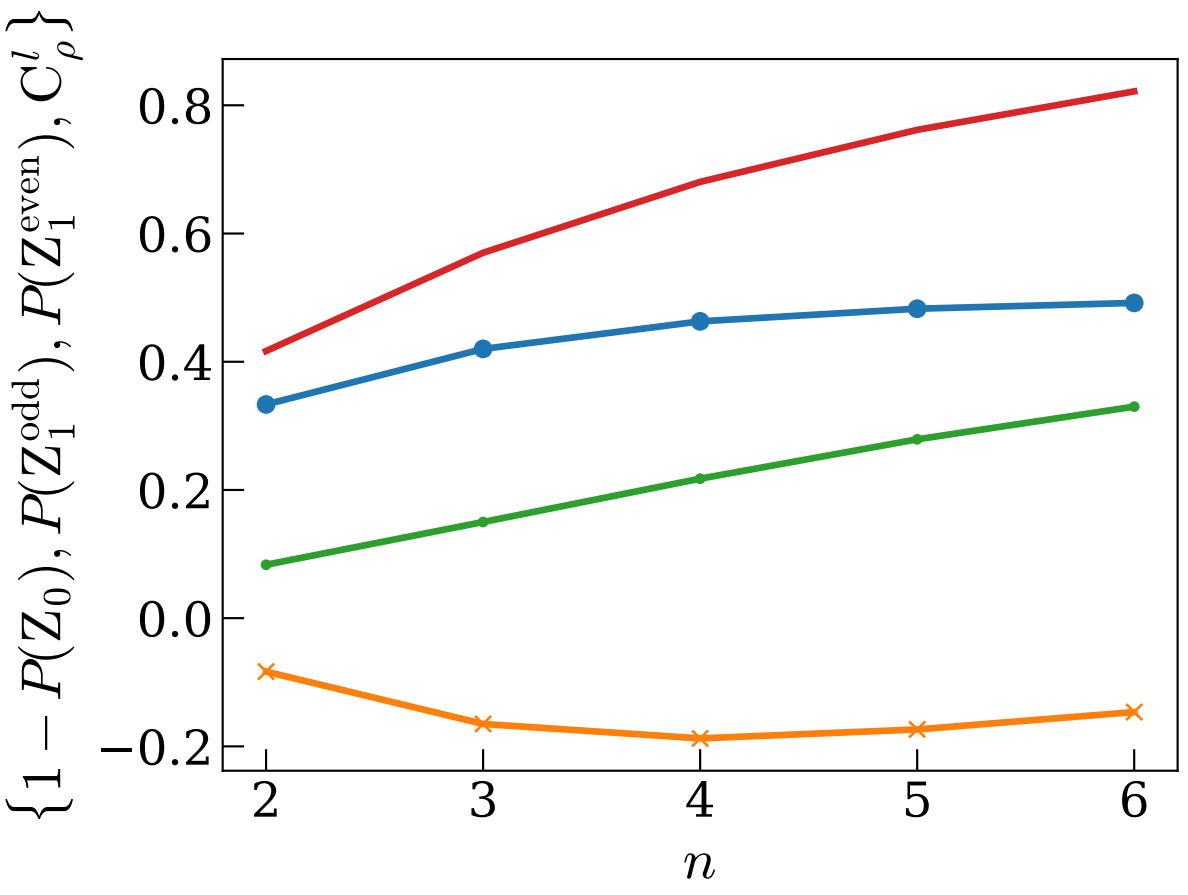

**Figure 4.** Test results for separable state $\rho = \rho' = \rho_{\text{GHZ}}(p_{\text{separable}})$. The red line is $1 - P(\mathcal{Z}_0)$, the blue circles are $P(\mathcal{Z}_1^{\text{odd}})$, the green dots represent $P(\mathcal{Z}_1^{\text{even}})$ and the yellow crosses are $\mathcal{C}^l_\rho$.

Let the two input states be two different mixed GHZ-like states such that $\rho = \rho_{\text{GHZ}}(p) = (1-p)|\text{GHZ}\rangle\langle\text{GHZ}| + p\frac{I_N}{N}$ and $\rho' = \rho_{\text{GHZ}}(q) = (1-q)|\text{GHZ}\rangle\langle\text{GHZ}| + q\frac{I_N}{N}$. These states give the results:

$$P(\mathcal{Z}_1^{\text{odd}}) = \frac{2^n - 1}{2^{n+1}}(p + q - pq) \tag{3.13}$$

$$P(\mathcal{Z}_1^{\text{even}}) = \frac{1}{2} - \frac{1}{2^n} - \left[\left(\frac{3}{4}\right)^n - \frac{3}{2^{n+1}}\right](p + q - pq) \tag{3.14}$$

$$1 - P(\mathcal{Z}_0) = \frac{1}{2} - \frac{1}{2^n} + \left[\frac{1}{2} + \frac{1}{2^n} - \left(\frac{3}{4}\right)^n\right](p + q - pq) \tag{3.15}$$

$$L = \frac{1}{2} - \frac{1}{2^n} - \left[\frac{1}{2} + \frac{1}{4^n} - \frac{3}{2^n} + \left(\frac{3}{4}\right)^n\right](p + q - pq)\,. \tag{3.16}$$

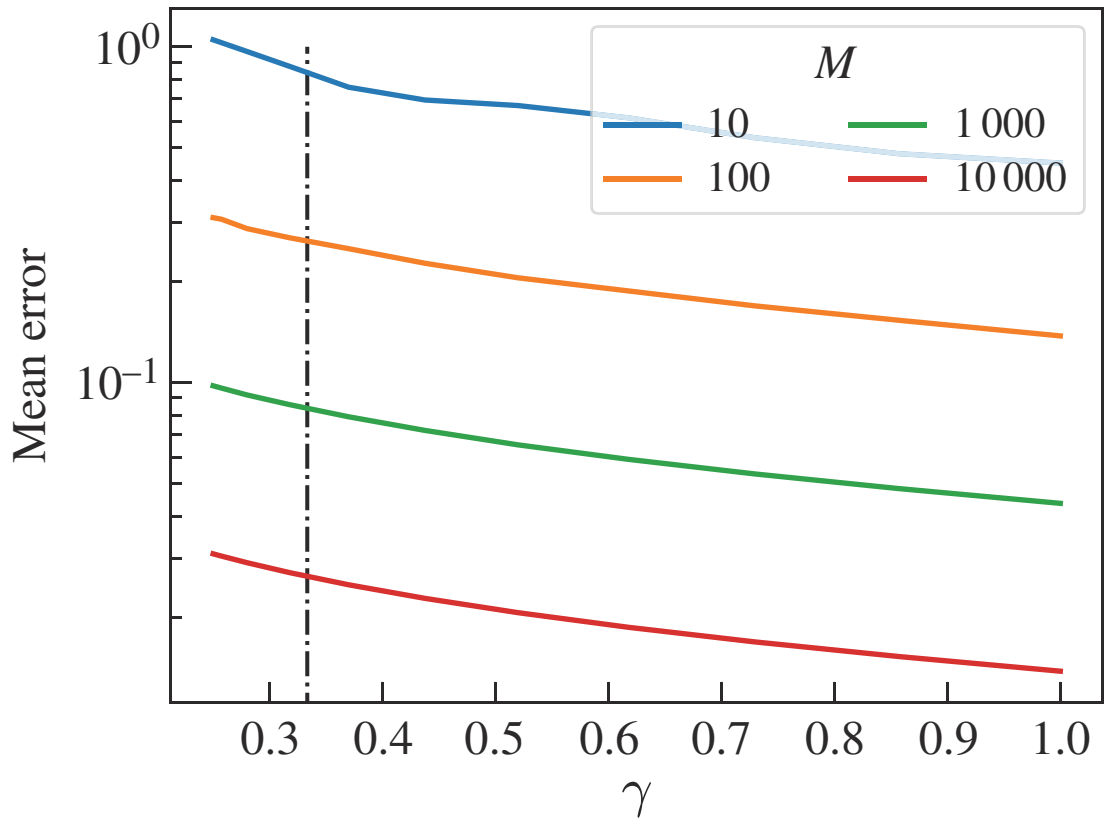

**Figure 5.** Mean error of $\widetilde{P}(\mathcal{Z}_1^{\text{even}})$ calculated from $M$ trials on $\rho_{\text{Bell}}(p) = (1-p)|\text{Bell}\rangle\langle\text{Bell}| + p\frac{I_N}{N}$, where $|\text{Bell}\rangle = \frac{1}{\sqrt{2}}(|00\rangle + |11\rangle)$ and $\gamma = 1 - \frac{3}{4}p(2-p)$. The dash-dot line shows $\gamma = \gamma_{\text{sep}}$. The mean error decreases with increased $M$.

The purity of $\rho$ is $\gamma = 1 - \frac{2^n-1}{2^n}p(2-p)$ and the purity of $\rho$, is $\gamma' = 1 - \frac{2^n-1}{2^n}q(2-q)$, and the fidelity of the two is $F = [\text{tr}\sqrt{\rho\rho'}]^2 = (\sqrt{1-p}\sqrt{1-q} + \sqrt{pq})^2$. The $n = 2$ case is shown in figure 6 for various values of $p$ and $\delta = q - p$.

Ideally, the values given by a non-identical input states anglement test should be some average over the values obtained from a test with input states $\rho$ &$\rho$, and from a test on input states $\rho'$ &$\rho'$, e.g. $L \approx \frac{1}{2}(\mathcal{C}_\rho^l + \mathcal{C}_{\rho'}^l)$. Let $\delta = q - p$. We have found that test results $X_i \in \{P(\mathcal{Z}_1^{\text{even}}), 1 - P(\mathcal{Z}_0), L\}$ obtained from input states $\rho = \rho_\psi(p)$ and $\rho' = \rho_\psi(q)$ relate to the mean average with

$$X_i(\rho, \rho') = \frac{1}{2}(X_i(\rho, \rho) + X_i(\rho', \rho')) - \frac{1}{2}\epsilon_i\delta^2, \tag{3.17}$$

where $|\epsilon_i| < 1$. The CE tests therefore estimate the average results of an ensemble to within $\frac{1}{2}\epsilon_i\delta^2$. For $X_{i=0} = P(\mathcal{Z}_1^{\text{even}})$ and $X_{i=2} = L$, $\epsilon_i$ is positive and for $X_{i=1} = 1 - P(\mathcal{Z}_0)$, $\epsilon_i$ is negative.

Therefore, $P(\mathcal{Z}_1^{\text{even}})$ underestimates the mean average value of $\frac{1}{2}(1 + \text{tr}[\rho^2]) - \frac{1}{2^n}\sum_{\alpha \in \mathcal{P}(S)} \text{tr}[\rho_\alpha^2]$ where $\rho = \rho_\psi(p)$ and $\rho = \rho_\psi(q)$ by $\frac{1}{2}\epsilon_0(q-p)^2$. However, recalling figure 4, $P(\mathcal{Z}_1^{\text{even}})$ overestimates the amount of entanglement in $\rho$ by $P(\mathcal{Z}_1^{\text{even}})$ when $\gamma = \gamma_{\text{separable}}$, since the CE should be zero for a separable state. To compare the underestimation from non-identical input states with the (maximum) overestimation from low purity, we show both in figure 7. If either of the input states have low purity, the resulting overestimation is much higher than the underestimation in all cases excepting $n = 2$ at high $\delta$. For high purity, with therefore very little expected overestimation, the average is underestimated by the relatively small value $\approx 0.1(q-p)^2$.

### (c) Upper bound on mixed state CE

Therefore in the expected experimental conditions of relatively small $\delta$, $P(\mathcal{Z}_1^{\text{even}})$ overestimates the amount of entanglement in input states $\rho$ and $\rho'$, and we therefore conclude that $P(\mathcal{Z}_1^{\text{even}})$ is a valid CE upper bound for mixed states and inequivalent input states. We define the bounds of total CE as

$$\begin{aligned}\mathcal{C}_\rho^u(S) &= \frac{1}{2}\left(1 + \text{tr}\left[\rho^2\right]\right) - \frac{1}{2^n}\sum_{\alpha \in \mathcal{P}(S)} \text{tr}\left[\rho_\alpha^2\right] \\ &\approx P(\mathcal{Z}_1^{\text{even}})\end{aligned} \tag{3.18}$$

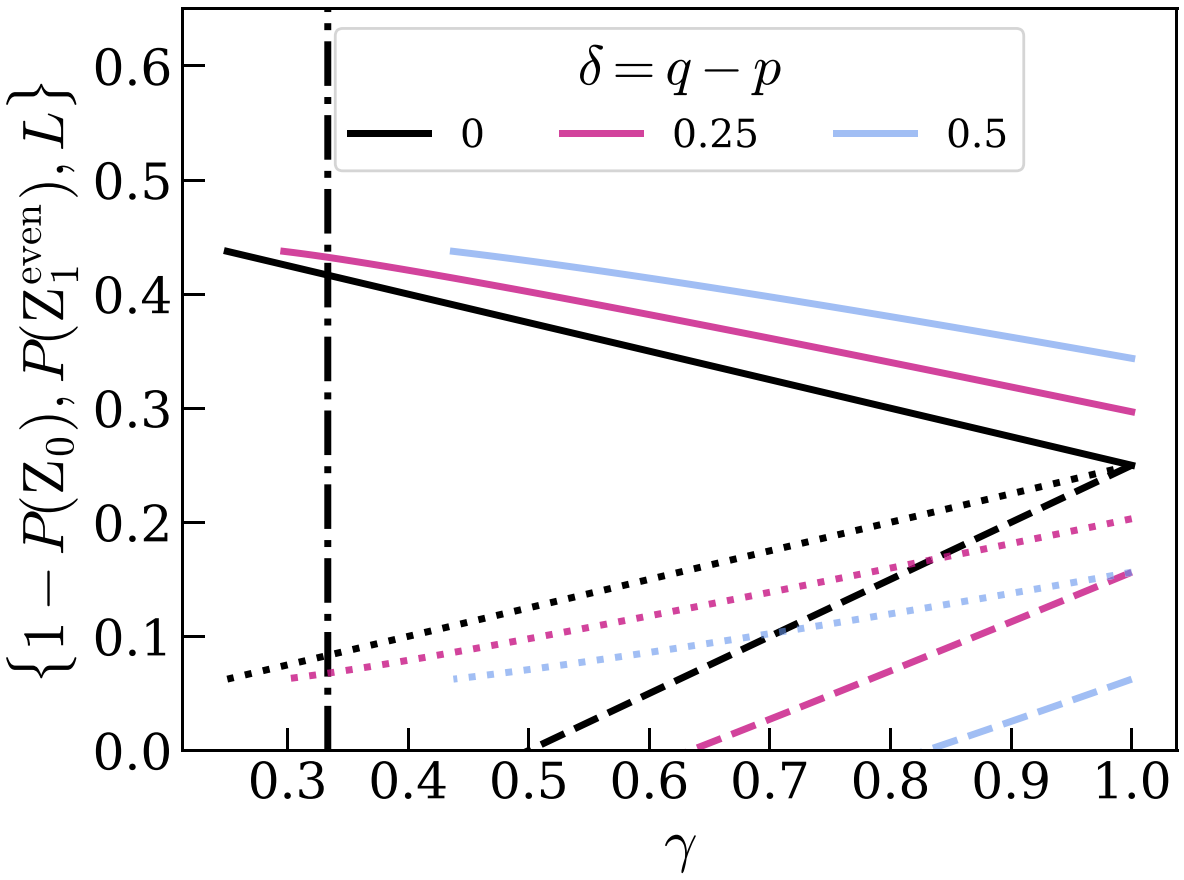

**Figure 6.** Test results for two different $n = 2$ Bell-like mixed input states $\rho = \rho_{\mathrm{Bell}}(p)$ and $\rho' = \rho_{\mathrm{Bell}}(q)$. The continuous line is $1 - P(\mathcal{Z}_0)$, the dotted line $P(\mathcal{Z}_1^{\mathrm{even}})$ and the dashed line $L = P(\mathcal{Z}_1^{\mathrm{even}}) - \left(1 - \frac{2}{2^n}\right) P(\mathcal{Z}_1^{\mathrm{odd}})$. Dash-dot lines show the value of $\gamma_{\mathrm{separable}}$ such that the state $\rho$ is separable for $\gamma \le \gamma_{\mathrm{separable}}$. $\delta = 0$ is represented by opaque lines, with decreasing opacity for increasing $\delta$.

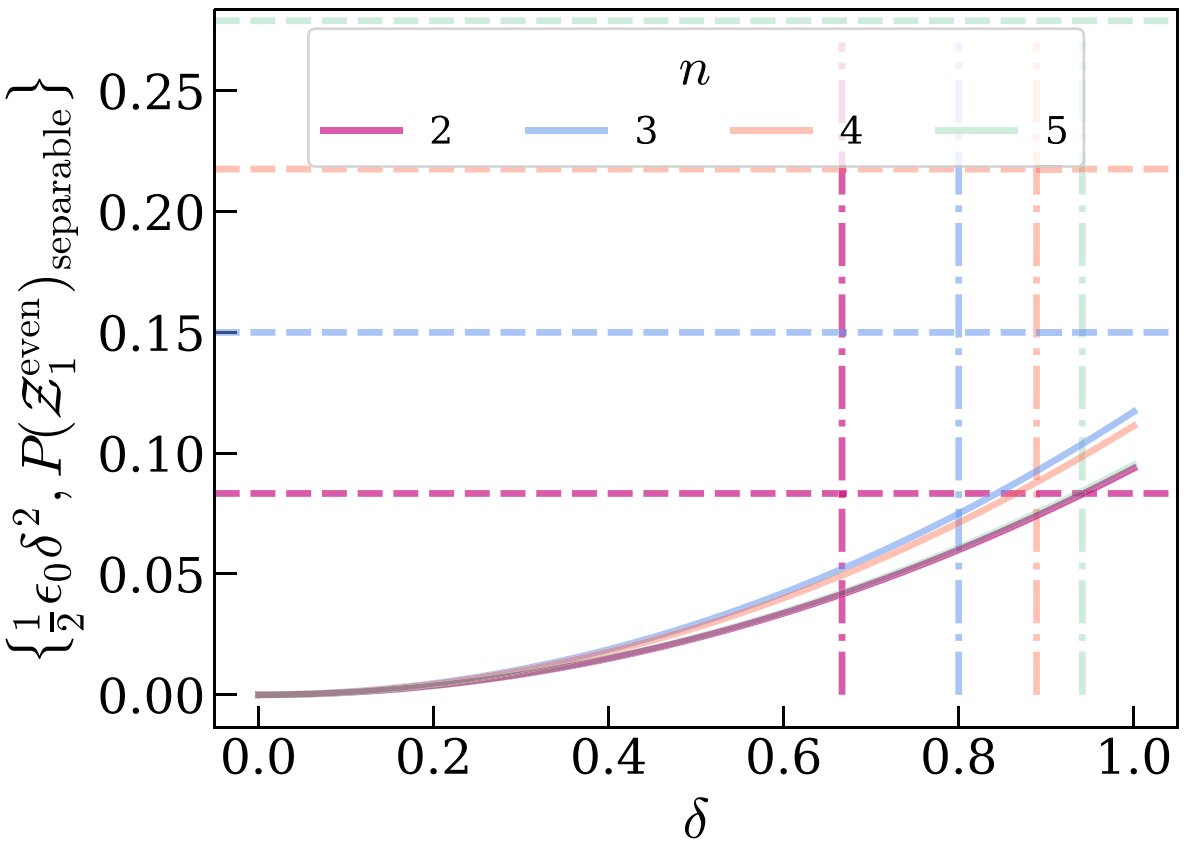

**Figure 7.** Solid lines show the underestimation $\frac{1}{2}\epsilon_P\delta^2 = P(\mathcal{Z}_1^{\mathrm{even}})(\rho, \rho') - \frac{1}{2}\left(P(\mathcal{Z}_1^{\mathrm{even}})(\rho, \rho) + P(\mathcal{Z}_1^{\mathrm{even}})(\rho', \rho')\right)$ for input states $\rho = \rho_{\mathrm{GHZ}}(p)$ and $\rho' = \rho_{\mathrm{GHZ}}(p + \delta)$. The dashed lines show maximum overestimation, which is equal to $P(\mathcal{Z}_1^{\mathrm{even}})$ with maximally mixed input states $\rho = \rho' = \rho_{\mathrm{GHZ}}(p_{\mathrm{separable}})$. Dash-dot lines show $\delta = p_{\mathrm{separable}}$. Increasing $n$ is represented by decreasing opacity.

$$\begin{aligned} \mathcal{C}_\rho^l(S) &= \frac{1}{2^n} + \left(1 - \frac{1}{2^n}\right)\mathrm{tr}\left[\rho^2\right] - \frac{1}{2^n}\sum_{\alpha \in \mathcal{P}(S)} \mathrm{tr}\left[\rho_\alpha^2\right] \\ &\approx P(\mathcal{Z}_1^{\mathrm{even}}) - \left(1 - \frac{2}{2^n}\right) P(\mathcal{Z}_1^{\mathrm{odd}}), \end{aligned} \tag{3.19}$$

where the approximates denote non-identical input states. The advantage of $\mathcal{C}_\rho^u = P(\mathcal{Z}_1^{\mathrm{even}})$ is that its operational meaning is identical to that of $\mathcal{C}_{|\psi\rangle}$: the 'concentration' of singlet Bell states detected consisting of subsystems in $\rho$. This is an overestimation of the entanglement present in mixed states however, as some of these singlet states represent the entanglement between the subsystems and the environment, aka mixedness.

Recall the maximum and minimum averages of a mixed state's CE over its decompositions, $\mathcal{C}_\rho^\cap$ and $\mathcal{C}_\rho^\cup$ from equations (3.5) and (3.6). The difference between these and our bounds is

$$\mathcal{C}_\rho^\cap - \mathcal{C}_\rho^u = \frac{1}{2}\left(1 - \mathrm{tr}\left[\rho^2\right]\right) - \frac{1}{2^n}\sum_{\alpha\in\mathcal{P}(S)}\left(\sup\sum_i p_i\,\mathrm{tr}\left[\rho_{\alpha,i}^2\right] - \mathrm{tr}\left[\rho_\alpha^2\right]\right), \tag{3.20}$$

$$\mathcal{C}_\rho^\cup - \mathcal{C}_\rho^l = \left(1 - \frac{1}{2^n}\right)\left(1 - \mathrm{tr}\left[\rho^2\right]\right) - \frac{1}{2^n}\sum_{\alpha\in\mathcal{P}(S)}\left(\inf\sum_i p_i\,\mathrm{tr}\left[\rho_{\alpha,i}^2\right] - \mathrm{tr}\left[\rho_\alpha^2\right]\right) \tag{3.21}$$

where $\rho_{\alpha,i} = \mathrm{Tr}_{S\setminus\alpha}|\psi_i\rangle\langle\psi_i|$. It can be seen from inspection that $\mathcal{C}_\rho^\cap - \mathcal{C}_\rho^u \leq \mathcal{C}_\rho^\cup - \mathcal{C}_\rho^l$, i.e. our new upper bound $\mathcal{C}_\rho^u$ is closer to the maximum average than the lower bound from prior work is to the minimum average.

Somewhat 'for free' (no additional quantum resources), the entanglement tests also estimate the average purity. For state $\rho$ with purity $\gamma$ and state $\rho'$ with purity $\gamma'$:

$$\begin{aligned} P(\mathcal{Z}_1^{\mathrm{odd}}) &= \frac{1}{2}\frac{2^n-1}{2^n}(p+q-pq) \\ &= \frac{1}{2}\left(1 - \frac{1}{2}(\gamma+\gamma')\right) - \frac{1}{4}\frac{2^n-1}{2^n}\delta^2. \end{aligned} \tag{3.22}$$

When $\rho = \rho'$, $P(\mathcal{Z}_1^{\mathrm{odd}}) = \frac{1}{2}(1-\gamma)$. We therefore define an estimate of the purity such that

$$\begin{aligned} \hat{\gamma} &= 1 - 2P(\mathcal{Z}_1^{\mathrm{odd}}) \\ &= \gamma - \frac{2^n-1}{2^n}(1-p)\delta \\ &= \gamma' + \frac{2^n-1}{2^n}(1-q)\delta\,. \end{aligned} \tag{3.23}$$

In conclusion, the c-SWAP and Bell basis entanglement tests can estimate the purity and the bounded CE of mixed qubit states to an acceptable accuracy, the considered sources of error being non-identical input states and the possibility of the state's CE being greater than the calculated upper bound. Mixed non-identical input states—the norm for experimental set-ups—are signified by non-zero $P(\mathcal{Z}_1^{\mathrm{odd}})$. The average purity of the input states can be estimated with $\hat{\gamma} = 1 - 2P(\mathcal{Z}_1^{\mathrm{odd}})$, which is $\epsilon \approx (1-p)|p-q|$ close to the actual purity $\gamma$, where $p$ and $q$ are Werner parameters. We have introduced an upper bound for the CE, estimable from an experiment with $\mathcal{C}_\rho^u = P(\mathcal{Z}_1^{\mathrm{even}})$. This upper bound is $\epsilon \approx \frac{1}{2}(p-q)^2$ lower than the average upper bound for non-identical ensembles, favouring the $\mathcal{C}_\rho^u$ of the least entangled of $\rho$ or $\rho'$. If $p$, $q$, or $|p-q| \gg 0$, overall $\mathcal{C}_\rho^u$ overestimates the amount of entanglement in $\rho$ and $\rho'$. $\mathcal{C}_\rho^u$ and $\mathcal{C}_\rho^l$ converge when $\rho$ is pure.

## 4. Concentratable entanglement for higher-dimensional states

We are interested in generalizing both the c-SWAP test and concentratable entanglement to experimentally relevant higher-dimensional states, specifically qudit states and coherent states. Specifically, we expect the CE to behave in ways found by previous entanglement measures. Higher dimensions allow the possibility for richer quantum architecture and simulation [29], simplified quantum circuits [30] and higher fault tolerance [31]. The use of the Bell measurement test on higher-dimensional states is uncertain as the test relies on the probability of measuring an antisymmetric Bell state [8]. However, there are no antisymmetric matrices with odd rank $D$ [47,48], so it is not clear how the Bell measurement test would extend to $D > 2$. Higher-dimensional states therefore require the c-SWAP version of the test.

## (a) Qudit states

### (i) Identical pure input states

Let $\rho = \rho' = |\psi\rangle\langle\psi|$, where $|\psi\rangle$ is a $D > 2$ dimensional qudit state. If using the c-SWAP test the control state remains a qubit ($D = 2$); therefore, the test's operation is unchanged, although the composite gate structure must be modified to achieve a SWAP operation on qudit states [49].

Although not explicitly stated, the proof of $1 - \frac{1}{2^n}\sum_{\alpha \in \mathcal{P}(S)} \mathrm{Tr}[\rho_\alpha^2] = 1 - P(\mathcal{Z}_0) = P(\mathcal{Z}_1^{\text{even}})$ for pure $\rho$ in Beckey *et al.* [6] is independent of dimension $D$. To find the effect of $D$ on CE, let us define a GHZ-like $n$-qudit state $|\psi\rangle = \sum_{j=0}^{D-1} A_j |j\rangle^n$, which gives

$$\mathcal{C}_{|\psi\rangle}(S) = P(\mathcal{Z}_1^{\text{even}}) = 4\left(\frac{1}{2} - \frac{1}{2^n}\right)\sum_{j=0}^{D-1}\sum_{k>j}^{D-1} |A_j^2 A_k^2| \,. \tag{4.1}$$

Figure 8 shows the CEs for maximally entangled 2-qudit states of dimension $D$ $\left(\mathcal{C}_{|\psi\rangle} = \frac{1}{2} - \frac{1}{2D}\right)$ along with those for maximally GHZ-entangled $n$-qubit states $\left(\mathcal{C}_{|\psi\rangle} = \frac{1}{2} - \frac{1}{2^n}\right)$ for comparison. Both increase with $D$ and $n$ respectively and tend to $\frac{1}{2}$; increasing $n$ has a greater effect on CE than increasing $D$, which is expected for an entanglement measure since the Hilbert space dimension for $n$ qubits is $2^n$.

### (ii) Identical mixed input states

As with qubit states, we model mixed qudit states with the $D$-dimensional Werner states $\rho = \rho' = (1 - p)|\psi\rangle\langle\psi| + p\frac{I_N}{N}$ where $N = D^n$. Let $|\psi\rangle_D$ be a $D$-dimensional Bell state $|\psi\rangle_D = \frac{1}{\sqrt{D}}\sum_{j=0}^{D-1} |jj\rangle$, therefore:

$$P(\mathcal{Z}_1^{\text{odd}}) = \frac{D^2 - 1}{2D^2} p(2 - p) \tag{4.2}$$

$$P(\mathcal{Z}_1^{\text{even}}) = \frac{1}{2} - \frac{1}{2D} - \frac{1}{4D^3}\left[D^2(D-2)p(2-p) + p(6D - (7D-8)p)\right] \tag{4.3}$$

shown in figure 9. Whereas mixed states of dimension $D = 2$ have CE in terms of $p(2 - p)$, higher-dimensional mixed states have an additional term $p(6D - (7D - 8)p)$. Unfortunately, $P(\mathcal{Z}_1^{\text{odd}}) \neq \frac{1}{2}(1 - \gamma)$ in all cases as $\gamma$ cannot be written in terms of $p(2 - p)$ for $D > 2$.

In conclusion, the c-SWAP test and CE can straightforwardly be applied to qudit states with equations (3.18) and (3.19). The terms $2^n$ are due to $|\mathcal{P}(S)| = 2^n$ and so these expressions are independent of dimension $D$, as can be intuited from the derivations [6,8] of $\mathcal{C}_{|\psi\rangle}$ and $\mathcal{C}_\rho^l$.

## (b) Entangled coherent states

We now consider how the c-SWAP test and CE can be applied to coherent states. Let $\rho = \rho' = |\psi\rangle\langle\psi|$ and therefore $\mathcal{C}_{|\psi\rangle}(S) = P(\mathcal{Z}_1^{\text{even}})$. Our proposed optical set-up to perform the c-SWAP test is shown in figure 10. The circuit is applied to the $k$-th mode in each of the input states $\rho$ and $\rho'$. This pair enters the circuit on spatial paths $A$ and $B$, respectively: the signals in one arm are then spatially swapped while the other arm experiences a phase-shift of $\frac{\pi}{2}$.

The phase-shift beam splitters can be represented by the transformation

$$\begin{bmatrix} E_3 \\ E_4 \end{bmatrix} = \frac{1}{\sqrt{2}}\begin{bmatrix} i & 1 \\ 1 & i \end{bmatrix}\begin{bmatrix} E_1 \\ E_2 \end{bmatrix}, \tag{4.4}$$

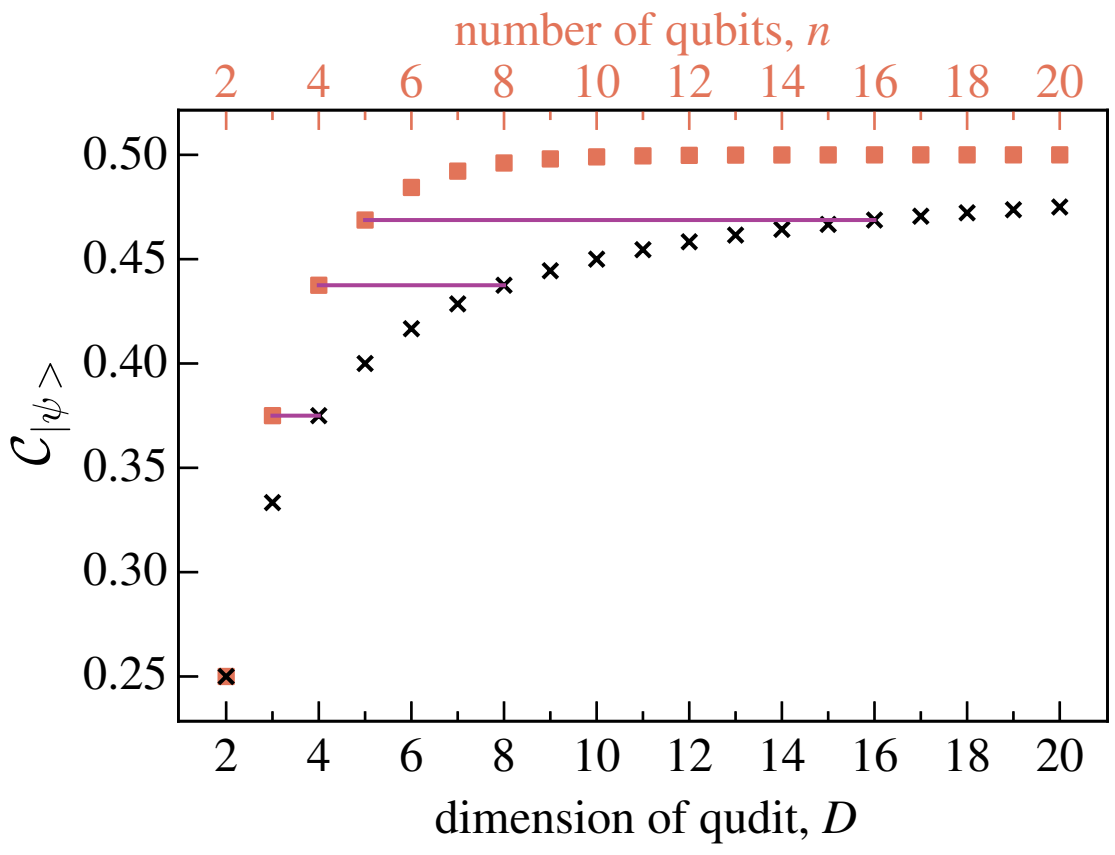

**Figure 8.** The total CE, $\mathcal{C}_{|\psi\rangle}$, given by the test for entanglement in maximally entangled 2-qudit input states $|\psi\rangle(n=2,D)=\frac{1}{\sqrt{D}}\sum_{k=0}^{D-1}|kk\rangle$ for various dimensions $D$ shown with black crosses. For comparison, the CE for maximally entangled $n$-qubit GHZ states $|\mathrm{GHZ}\rangle(n,D=2)=\frac{1}{\sqrt{2}}(|0\rangle^n+|1\rangle^n)$ are shown in orange squares. The horizontal purple lines show that the 2-qudit CE is related to the $n$-qubit CE with $\mathcal{C}_{|\mathrm{GHZ}\rangle(n=n',D=2)}=\mathcal{C}_{|\psi\rangle(n=2,D=2^{n'-1})}$.

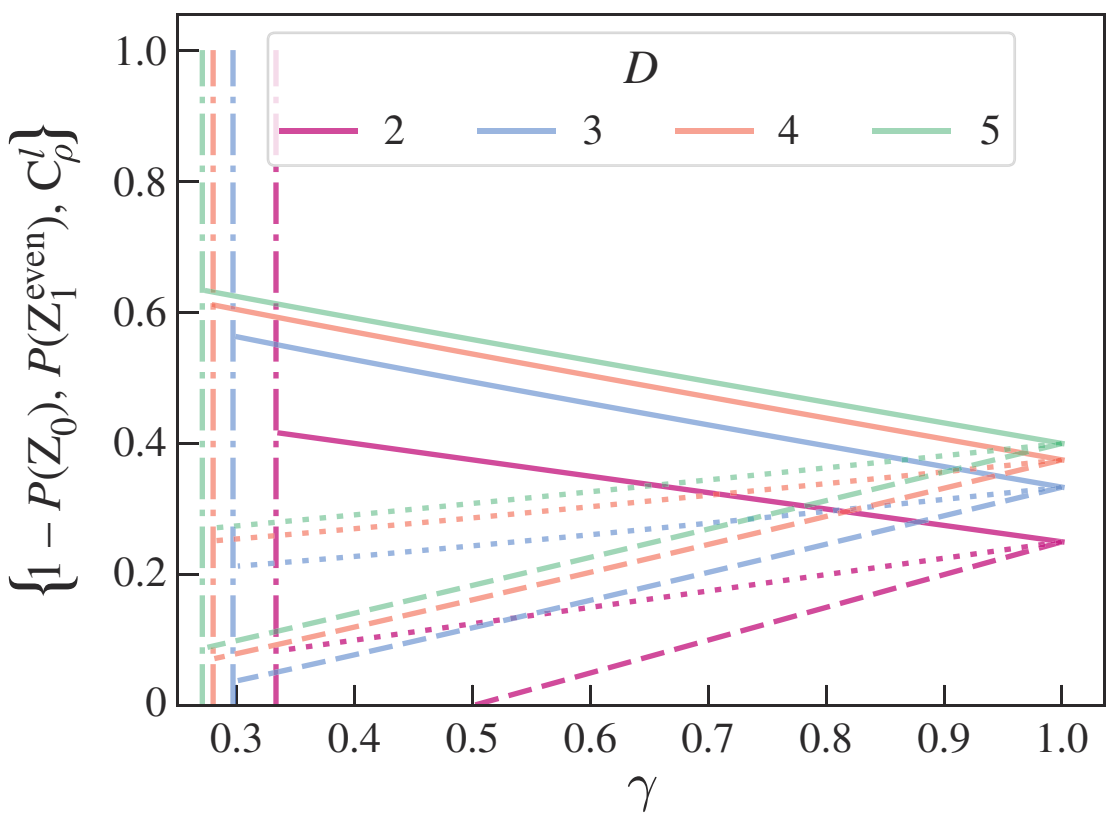

**Figure 9.** CE results for input states $\rho=\rho'=(1-p)|\psi\rangle\langle\psi|+p\frac{I_N}{N}$ where $N=D^n$ and $|\psi\rangle_D=\frac{1}{\sqrt{D}}\sum_{j=0}^{D-1}|jj\rangle$. Unbroken lines represent $1-P(\mathcal{Z}_0)$, dotted lines $P(\mathcal{Z}_1^{\mathrm{even}})$ and dashed lines $\mathcal{C}^l_{\rho_W}$. Dash-dot lines show the value of $\gamma_{\mathrm{separable}}$ such that the state is separable for $\gamma\le\gamma_{\mathrm{separable}}$.

where $E_1$ and $E_2$ are the input beams and $E_3$ and $E_4$ are the output beams. The action of the circuit is

$$\begin{bmatrix}E_AE'_B\\0\end{bmatrix}\xrightarrow{\text{PSBS1}}\frac{1}{\sqrt{2}}\begin{bmatrix}iE_AE'_B\\E_AE'_B\end{bmatrix}\xrightarrow{\text{SWAP}}\frac{1}{\sqrt{2}}\begin{bmatrix}iE_AE'_B\\E'_AE_B\end{bmatrix}\xrightarrow{\text{PSBS2}}\frac{1}{2}\begin{bmatrix}E'_AE_B-E_AE'_B\\iE_AE'_B+iE'_AE_B\end{bmatrix}.$$

The probability of a signal at detector C therefore is $\frac{1}{2}\mathrm{Tr}(E'_AE_B-E_AE'_B)$, which when $E=\rho_k$ is equivalent to the probability of measuring $|1\rangle_{C_k}$ in the c-SWAP test in figure 1. $P(\mathcal{Z}_1^{\mathrm{even}})$ is then the probability of detection (of anything other than the vacuum state) at C a non-zero even number of times.

The program [45] used to classically compute $P(\mathcal{Z}_1^{\mathrm{even}})$ requires the amplitudes of each mode in $|\psi\rangle$ to be able to be trivially swapped; therefore, we restrict our initial investigation to coherent state superpositions such as the two-mode entangled coherent states from equation (2.24). After a c-SWAP test on an identical ensemble of $|\psi\rangle = |\mathrm{ECS}_{\alpha,\beta}\rangle = \mathcal{N}_{\alpha,\beta}(A_{\alpha\alpha}|\alpha\rangle|\alpha\rangle + A_{\alpha\beta}|\alpha\rangle|\beta\rangle + A_{\beta\alpha}|\beta\rangle|\alpha\rangle + A_{\beta\beta}|\beta\rangle|\beta\rangle)$:

$$P(\mathcal{Z}_1^{\mathrm{even}}) = \mathcal{C}_{|\psi\rangle} = \mathcal{N}_{\alpha,\beta}^4\left(1 - \langle\alpha|\beta\rangle^2\right)^2 \left|A_{\alpha\alpha}A_{\beta\beta} - A_{\alpha\beta}A_{\beta\alpha}\right|^2, \tag{4.5}$$

which reduces to the two-qubit result if $\langle\alpha|\beta\rangle = 0$. To compare with entropy of entanglement, let $|\psi\rangle = (|\alpha\alpha\rangle + |\beta\beta\rangle)/\sqrt{1 + \langle\alpha|\beta\rangle^2}$. Therefore

$$\mathcal{C}_{|\psi\rangle} = \frac{1}{4}\frac{(1 - \langle\alpha|\beta\rangle^2)^2}{(1 + \langle\alpha|\beta\rangle^2)^2} \tag{4.6}$$

and

$$S_V(\rho_A) = \log(2) + \log\left(1 + \langle\alpha|\beta\rangle^2\right) - \frac{(1 - \langle\alpha|\beta\rangle)^2}{1 + \langle\alpha|\beta\rangle^2}\log(1 - \langle\alpha|\beta\rangle) - \frac{(1 + \langle\alpha|\beta\rangle)^2}{1 + \langle\alpha|\beta\rangle^2}\log(1 + \langle\alpha|\beta\rangle). \tag{4.7}$$

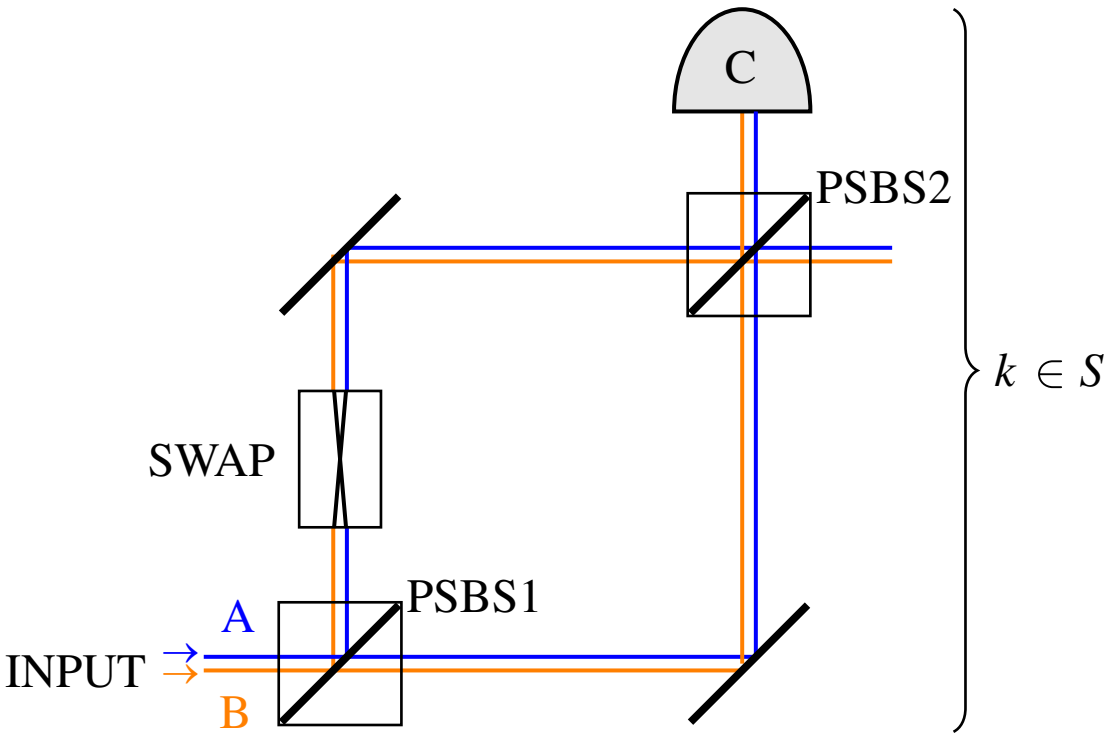

**Figure 10.** A proposed circuit, to be applied to each $k$-th mode, for implementing one round of the c-SWAP test on optical states. The $k$-th mode of the input states $\rho$ and $\rho'$ each enter the circuit on a different spatial path $A$ (blue) and $B$ (orange). PSBS1 and PSBS2 are phase-shift beam splitters, with transmitted beams experiencing a phase shift of $\frac{\pi}{2}$ and symmetric transmission/reflection probabilities. The SWAP operation crosses the two paths, such that $\{\rho_k\}_A\{\rho_k'\}_B \to \{\rho_k'\}_A\{\rho_k\}_B$. A detector is placed at $C$.

Figure 11 shows $\mathcal{C}_{|\psi\rangle}$ and the normalized entangled entropy $S_V(\rho_A)/4\log 2$ as a function of $\langle\alpha|\beta\rangle$. The shape of the two functions are very similar however $\mathcal{C}_{|\psi\rangle} < S_V(\rho_A)/4\log 2$ for all $\langle\alpha|\beta\rangle$. The closeness of the two measures' values leads us to conclude that they are quantifying the same property of $|\psi\rangle$ in a very similar way. The advantage of CE however is that it is analytically simpler.

To extend to $n$-mode ECSs, the $n$-qubit results from [7] are multiplied by $(1 - \langle\alpha|\beta\rangle^2)^{x \le n}$, where $x$ is the number of qubit swaps the state has undergone to give the amplitude in the final expression. For example, GHZ-like coherent states $|\psi\rangle = \mathcal{N}_{\mathrm{GHZ}}(A_\alpha|\alpha\rangle^n + A_\beta|\beta\rangle^n)$ where $\frac{1}{\mathcal{N}_{\mathrm{GHZ}}^2} = |A_\alpha|^2 + |A_\beta|^2 + (A_\alpha^* A_\beta + A_\beta^* A_\alpha)\langle\alpha|\beta\rangle^n$ give CE

$$\mathcal{C}_{|\psi\rangle} = \mathcal{N}_{\mathrm{GHZ}}^4\left(1 - \langle\alpha|\beta\rangle^2\right)^n\left(\frac{1}{2} - \frac{1}{2^n}\right)4|A_\alpha^2 A_\beta^2| \tag{4.8}$$

and W-like coherent states $|\psi\rangle = \mathcal{N}_{\mathrm{W}}\sum_{j=1}^n A_j|\beta\rangle^{j-1}|\alpha\rangle|\beta\rangle^{n-j}$ where $|\beta\rangle^0 = I_N$ and $\frac{1}{\mathcal{N}_{\mathrm{W}}^2} = \sum_{j=1}^{n}(|A_j|^2 + \langle\alpha|\beta\rangle^2 \sum_{k=1,k\neq j}^{n} A_j^* A_k)$ give

$$\mathcal{C}_{|\psi\rangle} = \mathcal{N}_{\mathrm{W}}^4\left(1 - \langle\alpha|\beta\rangle^2\right)^2 \sum_{j=1}^{n}\sum_{k>j}^{n} |A_j^2 A_k^2|. \tag{4.9}$$

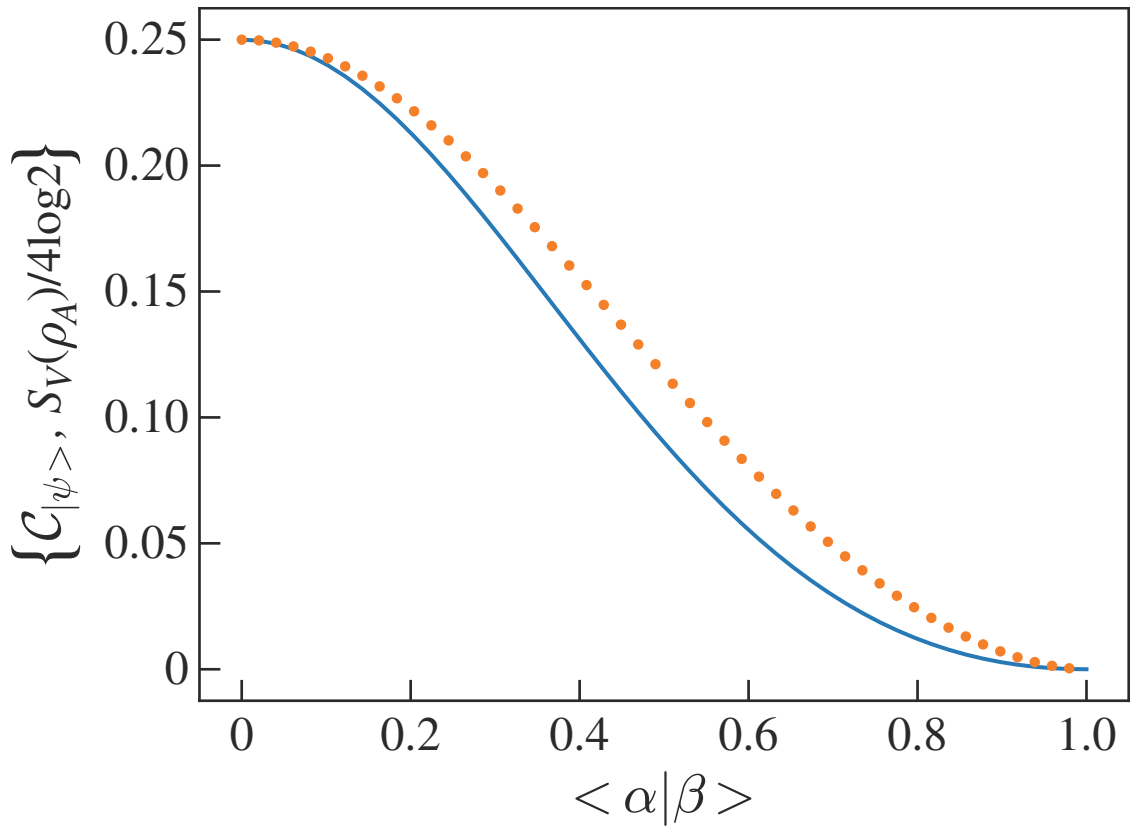

**Figure 11.** Graph showing the CE (blue solid line) and the normalized entanglement entropy (orange dotted line) of entangled coherent state $|\psi\rangle = (|\alpha\alpha\rangle + |\beta\beta\rangle)/\sqrt{1+\langle\alpha|\beta\rangle^2}$ for various $\langle\alpha|\beta\rangle$.

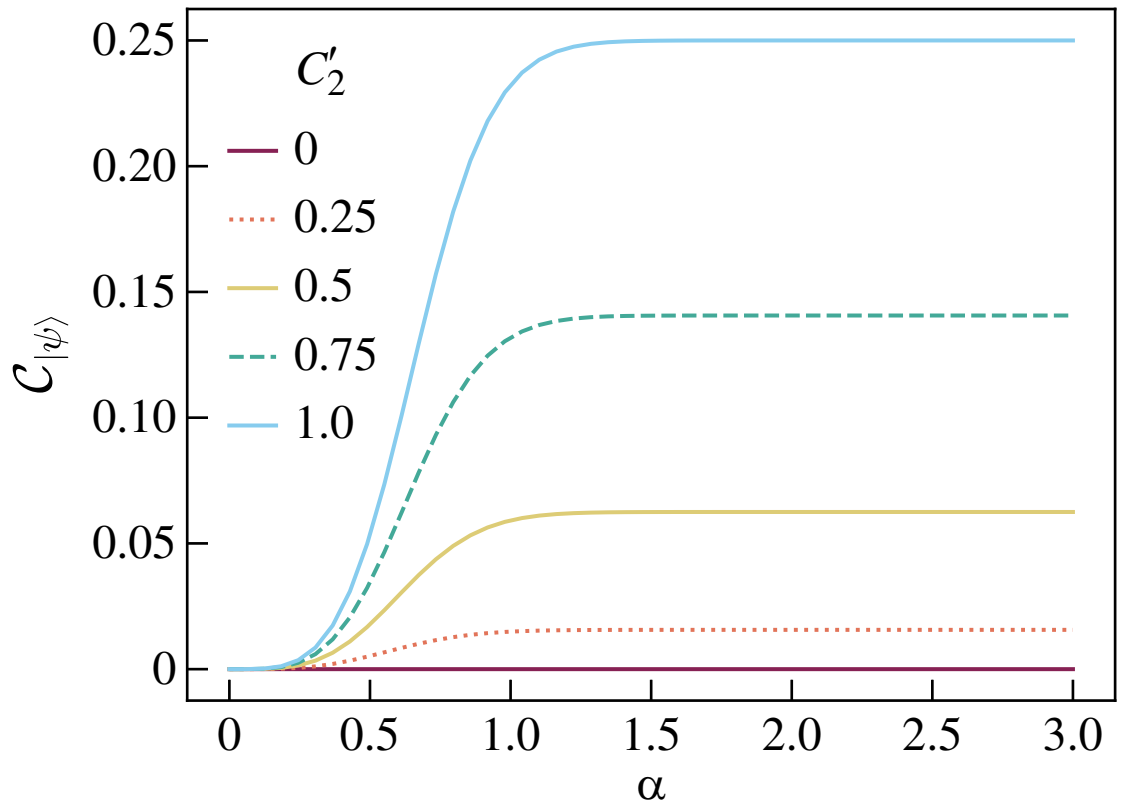

**Figure 12.** Graph showing $\mathcal{C}_{|\psi\rangle} = P(\mathcal{Z}_1^{\text{even}})$ for two-mode coherent state superpositions of the form given in (4.12), plotted against coherent state amplitude $\alpha$ for various values of $C_2'$ as defined in (4.14).

Further, this rule holds for mixed and non-identical input states. Therefore, previous favourable error analysis of qubit state results (found in §3 and Foulds *et al.* [7]) also hold for coherent states of a similar form.

Let us consider the specific example of

$$|\text{ECS}_{\alpha,-\alpha}\rangle = \mathcal{N}_{\alpha,-\alpha}(A_{++}|\alpha\rangle|\alpha\rangle + A_{+-}|\alpha\rangle|-\alpha\rangle + A_{-+}|-\alpha\rangle|\alpha\rangle + A_{--}|-\alpha\rangle|-\alpha\rangle)\,. \tag{4.10}$$

The c-SWAP test result for $|\psi\rangle = |\text{ECS}_{\alpha,-\alpha}\rangle$ is

$$P(\mathcal{Z}_1^{\text{even}}) = \mathcal{C}_{|\psi\rangle} = \frac{1}{4}C_2'^2\left(1 - e^{-4|\alpha|^2}\right)^2 \tag{4.11}$$

shown in figure 12, where

$$C_2' = 2\mathcal{N}^2_{\alpha,-\alpha}|A_{++}A_{--} - A_{+-}A_{-+}| \tag{4.12}$$

as an analogue to pure qubit state concurrence. The CE increases with $\alpha$ until $0.75 < \alpha < 1.5$ where it plateaus to the qubit state CE. This is expected as the greater $\alpha$, the smaller the overlap $\langle\alpha|-\alpha\rangle = \exp(-2|\alpha|^2)$ and the closer this overlap is to the qubit overlap $\langle 0|1\rangle = 0$. At small $\alpha$, the overlap is large and the closer the ECS to the product state $|\alpha\rangle|\alpha\rangle$.

Therefore the optical CE behaves as expected for an optical entanglement measure. However, since ECSs are less entangled than a qubit state of the same form, an entanglement test on an ECS will require more resources to estimate the smaller $P(\mathcal{Z}_1^{\text{even}})$. This is shown in figure 13, which shows the mean error of $\widetilde{P}(\mathcal{Z}_1^{\text{even}})$ against coherent state amplitude $\alpha$ for number of trails $M = 100$. When the $\alpha$ is large ($\alpha > 1$), the error is the same as for qubit states of the same form. When $\alpha = 0$, the error is unity as the probability to be estimated is zero. For $\alpha < \frac{1}{2}$ and\or $|A_{++}A_{--} - A_{+-}A_{-+}| < \frac{1}{2}$, the CE as estimated by the c-SWAP test has mean error $> 0.1$. The test is therefore intractable for small $\alpha$ coherent states.

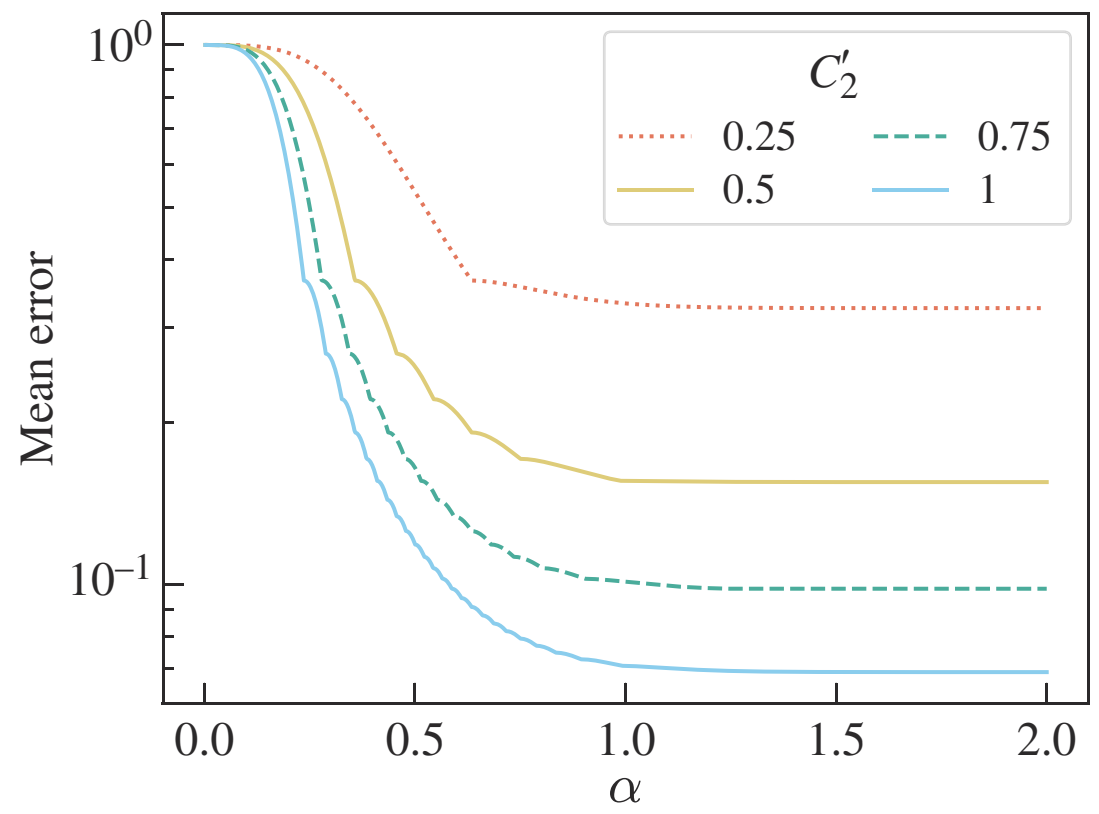

**Figure 13.** Mean error of $\widetilde{P}(\mathcal{Z}_1^{\text{even}})$ calculated from 100 trials on $|\text{ECS}_{\alpha,-\alpha}\rangle$ from equation 4.10, plotted against coherent state amplitude $\alpha$ for various values of $C_2'$ as defined in equation (4.14).

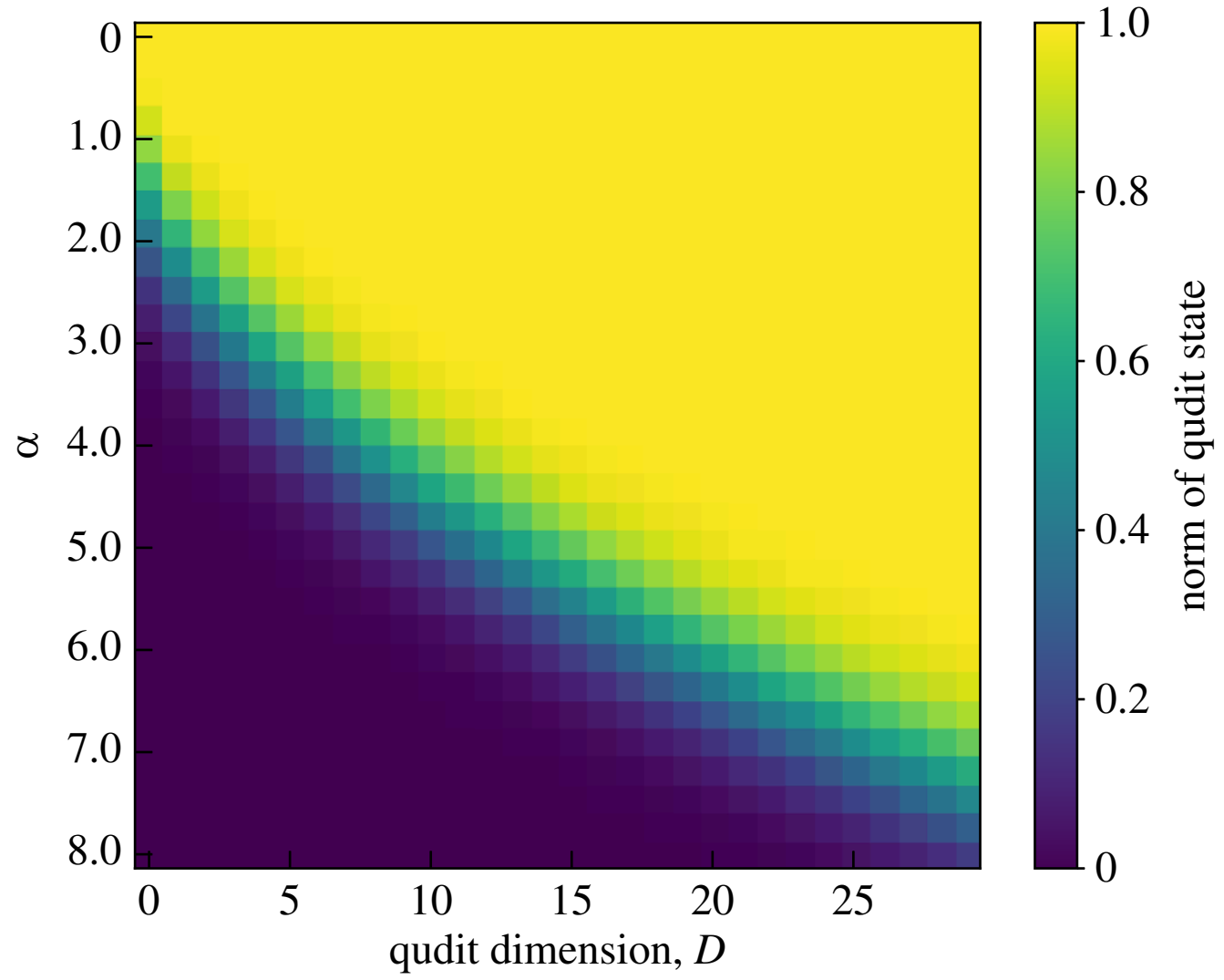

**Figure 14.** The norm of the qudit approximated ECS, equation (4.17), as a heatmap across different qudit dimensions and coherent state amplitudes. This was used to determine a value of $D$ for which the qudit state accurately models the ECS, equation (4.16), across the coherent state amplitude $\alpha$ range considered.

Also, note that the above scheme is very similar to the comparison of coherent states [50], which only requires one beam splitter per circuit. However, we have seen that the coherent state CE is in terms of $(1 - \langle\alpha|\beta\rangle^2)$, whereas the comparison scheme gives probabilities in terms of $(1 - \langle\alpha|\beta\rangle)$. Further work should be done to investigate whether the circuit given in figure 10 could be simplified.

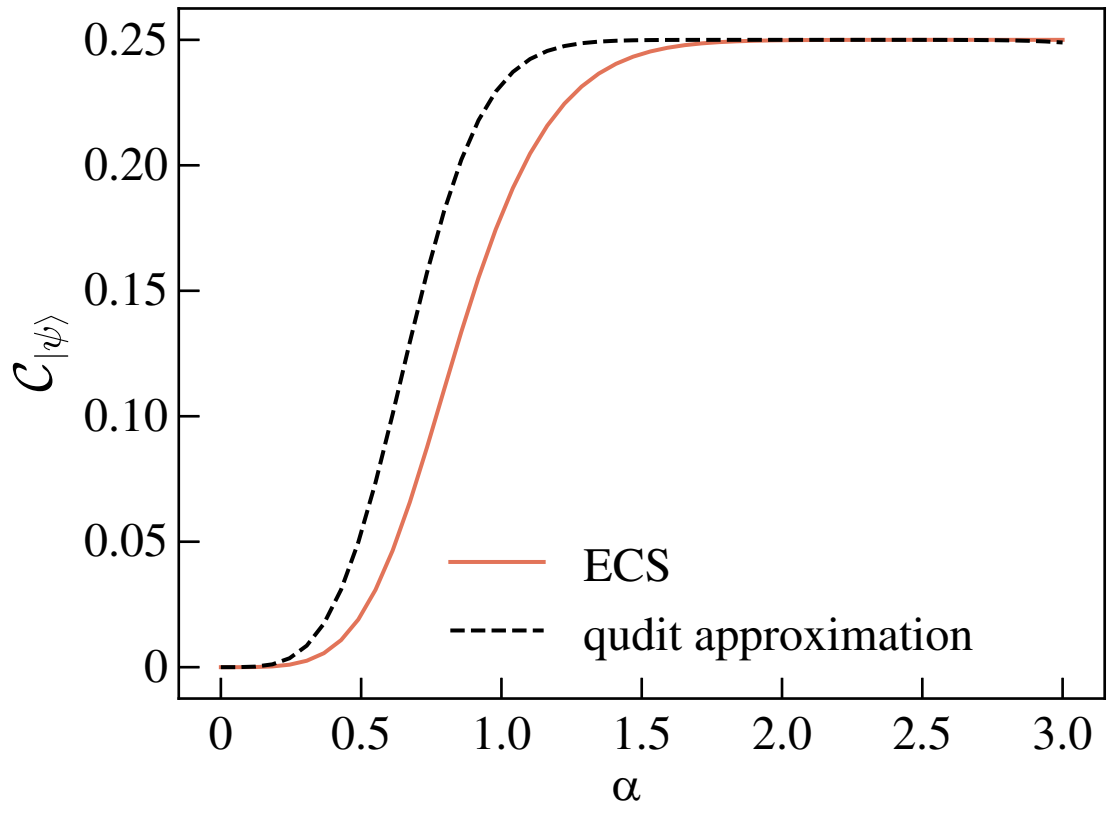

**Figure 15.** A comparison of $\mathcal{C}_{|\psi\rangle} = P(\mathcal{Z}_1^{\text{even}}))$ for ECS equation (4.16) and a qudit approximation ECS equation (4.17) against coherent state amplitude $\alpha$.

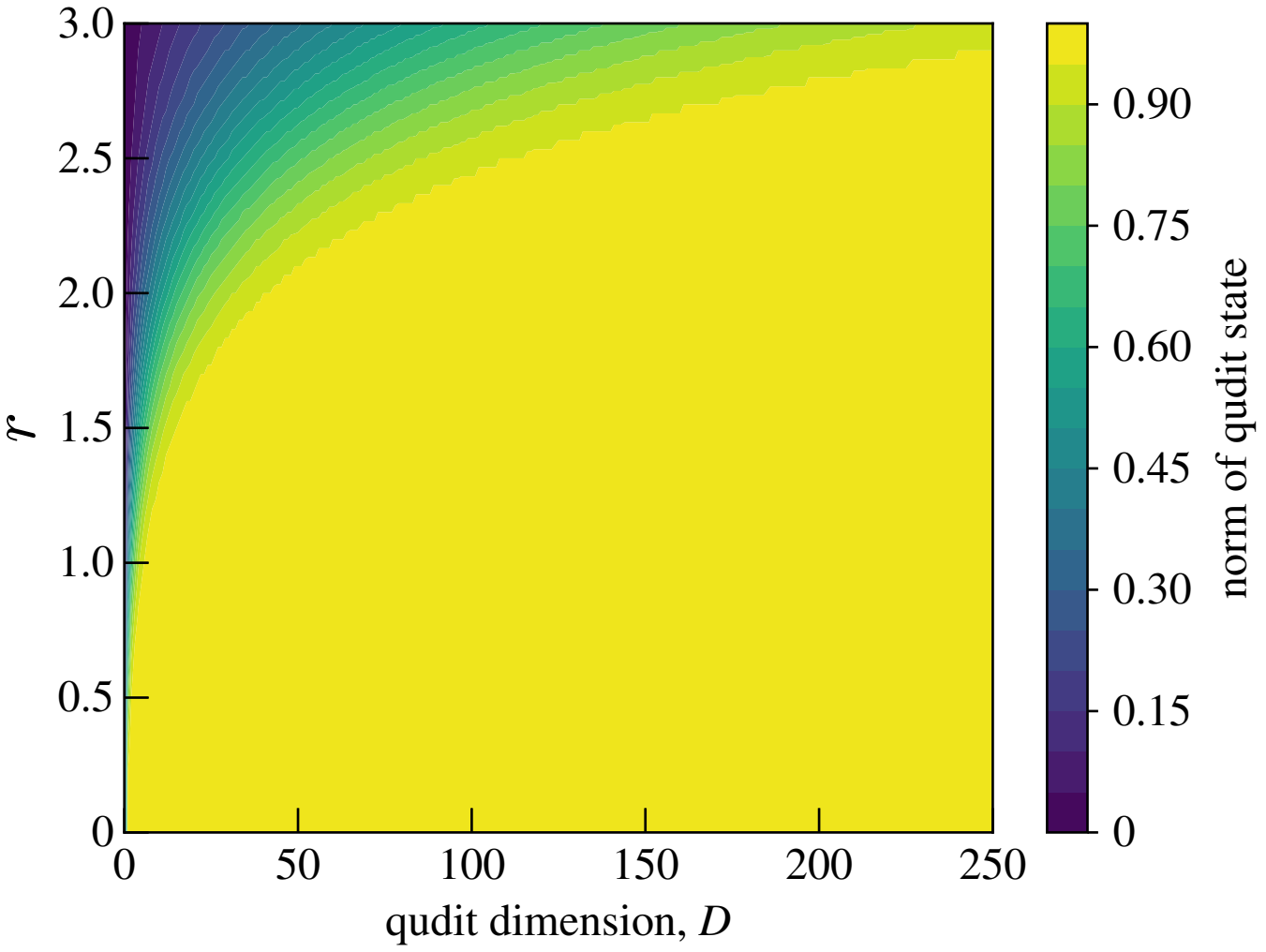

**Figure 16.** The norm of the approximated TMSV state equation (4.18) for various qudit dimension $D$ and squeeze parameter $r$.

## (c) Coherent states as high dimensional qudits

We wish to calculate the CE of optical states not supported by the existing code. An alternative method when considering optical states is to approximate them with high dimensional qudits, with the $D$-dimensional approximation:

$$\left|\alpha_{\text{qudit}}\right\rangle = e^{-\frac{|\alpha|^2}{2}} \sum_{j=0}^{D-1} \frac{\alpha^j}{\sqrt{j!}} |j\rangle. \tag{4.13}$$

The c-SWAP test as shown in figure 1*a* can then be applied to this state.

Consider the simple ECS

$$|\psi_\alpha\rangle = \mathcal{N}_{\alpha,-\alpha}(|\alpha\rangle|\alpha\rangle + |-\alpha\rangle|-\alpha\rangle), \tag{4.14}$$

which can be approximated by

$$\left|\psi_{\text{qudit}}\right\rangle = e^{-|\alpha|^2} \sum_{j,k=0}^{14} (1 + (-1)^{j+k}) \frac{\alpha^j}{\sqrt{j!}} \frac{\alpha^k}{\sqrt{k!}} |j\rangle|k\rangle, \tag{4.15}$$

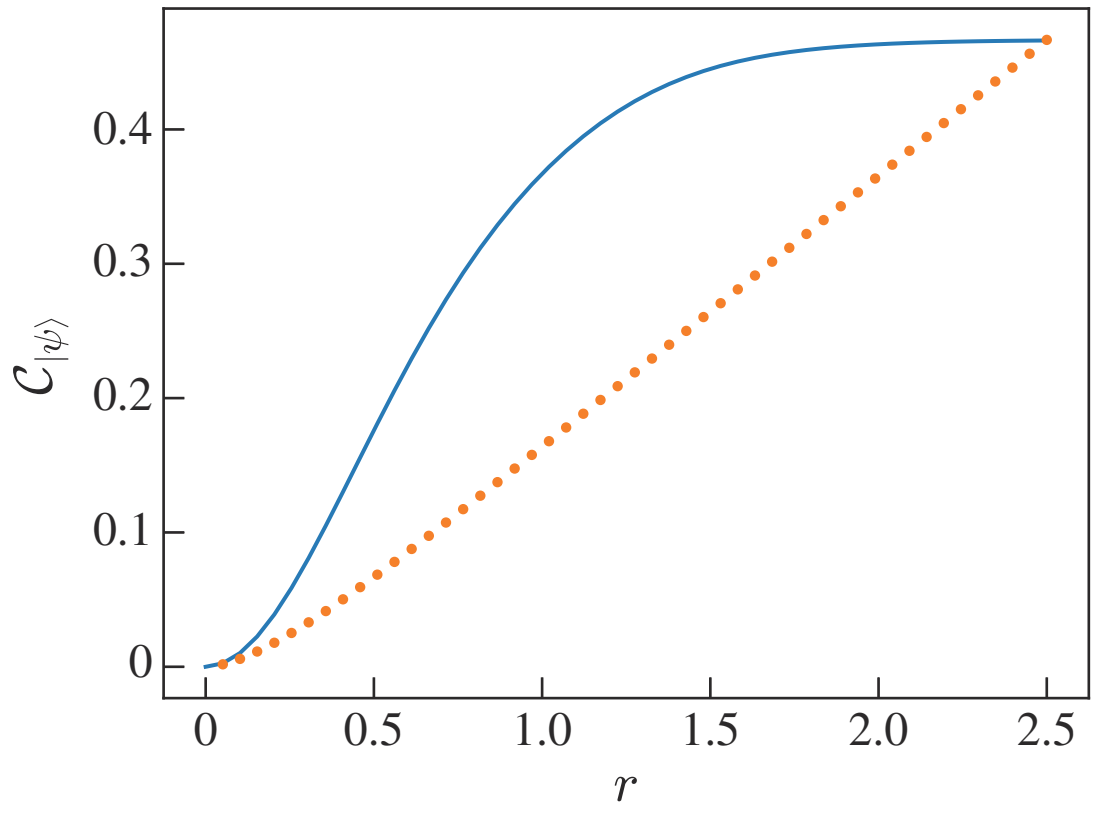

**Figure 17.** CE of a two-mode squeezed vacuum (TMSV) state approximated by a $D = 250$ qudit state from equation (4.18) (solid blue line) alongside $\frac{S_V(\rho_A)(r)}{2S_V(\rho_A)(r=3)}$ of the TMSV state (dotted orange line).

where we have chosen $D = 15$ so that $|\psi_{\text{qudit}}\rangle$ is approximately normalized in the range $0 < \alpha < 3$ (see figure 14). The c-SWAP test output probabilities against $\alpha$ for this state are shown in figure 15. In the same plot are the results for $|\psi_\alpha\rangle$ for comparison.

The behaviour of $P(\mathcal{Z}_1^{\text{even}})$ for each state are very similar, tending to the same value, but the qudit approximation overestimates in the region of $0 < \alpha < 1.8$.

Now consider an optical state that the c-SWAP test cannot be trivially applied to: the two-mode squeezed vacuum state $|\text{TMSV}_\alpha\rangle = S_2(\xi)|0,0\rangle$, where $S_2(\xi)$ is the two-mode squeeze operator defined in equation (2.22). These states have been used to demonstrate the EPR paradox experiment with continuous position and momentum variables [51].

The qudit approximation of a TMSV state where $0 \le \alpha \le 3$ is [52]

$$|\text{TMSV}_{\text{qudit}}\rangle = \frac{1}{\cosh r}\sum_{j=0}^{249}(-e^{i\theta}\tanh r)^j|jj\rangle, \tag{4.16}$$

the normalization values of which are shown in figure 16. Unfortunately, this state requires a very high dimension, $D = 250$, and therefore would likely be intractable in an experimental setting. Regardless, the CE test probability result is

$$\mathcal{C}_{|\text{TMSV}_{\text{qudit}}\rangle} = \frac{1}{2\cosh^4 r}\sum_{j=0}^{249}\sum_{k=0,k\neq j}^{249}(\tanh r)^{2(j+k)}. \tag{4.17}$$

This is shown in figure 17, alongside the normalized entanglement entropy $\frac{S_V(\rho_A)(r)}{2S_V(\rho_A)(r=3)}$, where $S_V(\rho_A)(r) = \cosh^2 r \log \cosh^2 r - \sinh^2 r \log \sinh^2 r$ [35]. Unfortunately, $\mathcal{C}_{|\text{TMSV}_{\text{qudit}}\rangle}$ does not seem to be a good estimation of $\mathcal{C}_{|\text{TMSV}_\alpha\rangle}$. The entanglement entropy increases with $r$ linearly and indefinitely; however, the CE of the qudit approximation has a maximum at $r \approx 2.5$, beyond which the CE decreases. This is due to the qudit state CE term $\frac{1}{2} - \frac{1}{D}$, which tends to $\frac{1}{2}$ as $D$ increases. This suggests CE tests on qudit approximations of coherent states are not suitable for squeezed states.

Since the CE of the ECS qudit approximation is close to the CE of ECSs (and the CE of ECSs are close to the entanglement entropy), the disparity in the TMSV results may point to an inherent limitation in the entanglement tests to estimate a TMSV state's degree of entanglement.

The c-SWAP test and therefore concentratable entanglements can be immediately applied to qudit states and entangled coherent states, however further work is needed to assess its suitability for squeezed states. In general, CE increases with dimension $D$ and coherent amplitude $\alpha$, strengthening it as a multi-dimensional entanglement measure.

## 5. Conclusions

In conclusion, we have built on past work that defines the CE of pure states [6] and the lower bound on the CE of mixed states [8] by defining the upper bound of the (total) CE of $n$-party mixed states as

$$\begin{aligned} \mathcal{C}^u_\rho(S) &= \frac{1}{2}\left(1 + \mathrm{tr}\left[\rho^2\right]\right) - \frac{1}{2^n}\sum_{\alpha \in \mathcal{P}(S)} \mathrm{tr}\left[\rho_\alpha^2\right] \\ &\approx P(\mathcal{Z}_1^{\text{even}})\,. \end{aligned} \tag{5.1}$$

The upper and lower bounds converge when all input states are identical and pure. The upper and lower bounds are the tightest experimentally obtainable values that bound zero when $\rho$ is separable. Furthermore we have shown that when the input states are not identical, these bounds give the average CE of the input ensemble to within $\frac{1}{2}\delta^2$, where $\delta$ is the difference between two Werner states' parameters. Further work should generalize these bounds to subsystem CE.

$\tilde{\mathcal{C}}^u_\rho$ and $\tilde{\mathcal{C}}^l_\rho$, where $\rho$ is qubit, can be estimated from $2M$ copies of $\rho$ using the Bell basis test [8]. The mean error of $\tilde{\mathcal{C}}^u_\rho$ is logarithmically dependent on the purity of $\rho$.

In addition, we have expanded the definition of the CE and its corresponding test to higher dimensions. The c-SWAP test for entanglement applied to higher-dimensional states cannot estimate purity as with the two-dimensional case, however the CE behaves as expected for a higher-dimensional entanglement measure. We considered the test applied to entangled coherent states; the CE of these behave similarly to the entropy of entanglement but with much simpler analytical expressions, therefore we believe the CE is the more operationally friendly measure of entanglement. However, estimating the CE of states with small coherent state amplitudes via the c-SWAP test is intractable due to the relatively low level of entanglement inherent to such states. Further work is required to assess the validity of the CE of other optical states such as squeezed states and OAM states.

Data accessibility. Code available at [45].

Declaration of AI use. We have not used AI-assisted technologies in creating this article.

Authors' contributions. S.F.: data curation, formal analysis, investigation, software, supervision, writing—original draft, writing—review and editing; O.P.: data curation, formal analysis, investigation, software, writing—original draft; V.K.: conceptualization, funding acquisition, project administration, resources, supervision, writing—review and editing.

All authors gave final approval for publication and agreed to be held accountable for the work performed therein.

Conflict of interest declaration. We declare we have no competing interests.

Funding. S.F. was supported by a UK EPSRC funded DTG studentship project reference 2210204.

Acknowledgements. Thank you to Tim Spiller for introducing us to the c- SWAP test and many excellent conversations. Thanks to Gerard Pelegrí for code and assistance in creating figures 5 and figure 13.